\documentclass[prm,reprint]{revtex4-2}
\usepackage[utf8]{inputenc}
\usepackage[T1]{fontenc}

\usepackage{setspace, parskip, xcolor, breakcites, hyphenat, graphicx, epstopdf, amsmath, amssymb, amsfonts, array,
	verbatim, url, subcaption, booktabs, multirow, etoolbox,siunitx, etoolbox, tikz,stackengine, textcomp,
	gensymb, siunitx,tabularx,array, soul}

\newcolumntype{H}{>{\setbox0=\hbox\bgroup}c<{\egroup}@{}}
\usepackage[toc,page,titletoc]{appendix}
\usepackage[inline]{enumitem}
\usepackage[version=4]{mhchem}
\usepackage[T1]{fontenc}
\usepackage[font=small,labelfont=bf]{caption}
\usetikzlibrary{shadings}
\definecolor{srm}{HTML}{034da1}
\definecolor{cof}{RGB}{219,144,71}
\definecolor{pur}{RGB}{186,146,162}
\definecolor{greeo}{RGB}{91,173,69}
\definecolor{greet}{RGB}{52,111,72}
\definecolor{mplc0}{HTML}{1f77b4}
\definecolor{mplc1}{HTML}{ff7f0e}
\definecolor{mplc2}{HTML}{2e7d32}
\definecolor{mplc3}{HTML}{d32f2f}
\definecolor{mplc4}{HTML}{9467bd}
\definecolor{mplc5}{HTML}{8c564b}
\definecolor{mplc6}{HTML}{e377c2}
\definecolor{mplc7}{HTML}{7f7f7f}
\definecolor{mplc8}{HTML}{bcbd22}
\definecolor{mplc9}{HTML}{17becf}
\definecolor{Astral}{HTML}{1F77B4}
\definecolor{BG80}{HTML}{37474f}
\definecolor{Green}{HTML}{4CA540}
\definecolor{Red}{HTML}{F44336}
\usepackage[colorlinks=true, linkcolor=mplc3, urlcolor=mplc3, filecolor=mplc0, citecolor=mplc2,
	pdfstartview=FitV, pdftitle={}, pdfauthor={}, pdfsubject={}, pdfkeywords={}, pdfpagemode={},
	bookmarksopen=true, breaklinks]{hyperref}

\usepackage[capitalise,nameinlink]{cleveref}
\crefname{supp}{Supplement}{Supplements}
\PassOptionsToPackage{hyphens}{url}

\newcommand{\figref}[1]{figure (\ref{#1})}

\newcommand{\tabref}[1]{table (\ref{#1})}

\date{\today}

\makeatletter
\def\@email#1#2{%
	\endgroup
	\patchcmd{\titleblock@produce}
	{\frontmatter@RRAPformat}
	{\frontmatter@RRAPformat{\produce@RRAP{*#1\href{mailto:#2}{#2}}}\frontmatter@RRAPformat}
	{}{}
}%
\makeatother
\begin{document}

\title{Beyond Diamond: Interpretable Machine Learning Reveals Design Principles for Quantum Defect Host Materials}
\author{Mohammed Mahshook}
\author{Rudra Banerjee*}\email{rudrab@srmist.edu.in}
\affiliation{Department of Physics and Nanotechnology, SRM Institute of Science and Technology, Kattankulathur, Tamil
	Nadu, 603203, India}

\begin{abstract}
	Solid-state spin defects in wide-bandgap semiconductors are leading candidates for quantum information processing, but systematic
	identification of suitable host materials remains limited by the cost of first-principles screening across vast chemical spaces.
	We address this with a composition-only machine learning framework built on heterogeneous Rashomon set ensembles: by contrasting
	the feature attributions of seven diverse classifiers, we extract consensus design rules that no single model identifies
	alone---filled valence $s$-, $d$-, and $f$-shells, low chemical heterogeneity, and enrichment in C, S, Si, and O favor quantum
	compatibility.  Screening $\approx$45,000 thermodynamically stable compounds, we identify 122 high-confidence candidates
	(confidence $>0.95$), recovering most experimentally verified hosts (C, \ce{SiC}, \ce{ZnO}, \ce{ZnS}) and predicting unexplored
	materials including \ce{TiO2}, \ce{PbWO4}, and layered chalcogenides (\ce{HfS2}, \ce{ZrS2}).  Density functional perturbation
	theory calculations on 12 representative materials validate dielectric screening as a coherence proxy ($R^2 = 0.89$ against
	experimental $T_2$), and vacancy calculations for \ce{TiO2} reveal deep, isolated mid-gap states favorable for spin-defect
	hosting.  The framework provides transferable, physically grounded design principles for rational quantum materials discovery
	beyond traditional carbide and nitride hosts.
\end{abstract}
\maketitle

\section{Introduction}
Quantum Information Science is revolutionizing computing, communication, and sensing through advances in qubit
platforms\cite{I21QuantumReadiness, divincenzo2025thirty}.  Despite this progress, realization of large-scale systems such as a
fault-tolerant quantum computer is still challenging due to stringent hardware requirements: reliable initialization, universal
control, robust readout, and long spin coherence times\cite{I2QCReview, I3QCReview}. Current platforms such as ultracold
atoms\cite{kaur2025neutral}, superconducting circuits\cite{I6SQubits}, and photonic systems\cite{I6PQubits} show promise but
struggle with maintaining coherence and achieving fault-tolerant integration\cite{IE5}.

Solid-state spin defects in wide-bandgap semiconductors offer a compelling alternative, combining atomic clock-like coherence
with potential for device-level integration across operating temperatures\cite{awschalom_quantum_2018, seo2017designing}. The
nitrogen-vacancy (NV$^-$) center in diamond exemplifies this approach, featuring optically addressable spin states and
exceptional room-temperature coherence\cite{I14NVCenter, I15NVCenter}. However, diamond faces practical constraints: synthesis
difficulty, limited tunability, and poor compatibility with low-dimensional and integrated device
architectures\cite{Atature_2018, wolfowicz_quantum_2021}.

To overcome these limitations, researchers have pursued NV-like defects in alternative hosts such as silicon carbide
(\ce{SiC})\cite{I15SiCQubits}, hexagonal boron nitride (\ce{hBN})\cite{carbone2025quantifying}, and transition metal
dichalcogenides (TMDs)\cite{I10, I16TMDQubits}. Yet identifying suitable host-defect combinations remains difficult. Ideal
quantum defects require well-isolated mid-gap states, optical addressability, long spin coherence, structural stability, and
realistic synthesis feasibility\cite{M1, IE8, I17}. This challenge is compounded by the vast materials landscape: even under
strict chemical constraints, millions of stoichiometric compounds exist with multiple polymorphs, each exhibiting different
quantum properties\cite{I12, IE8}.

Essential host attributes like wide bandgaps, low nuclear spin abundance, high dielectric screening, and non-polar symmetry are
rare and highly sensitive to composition and crystal structure\cite{Seo_2016, ivady_ab_2019}. First-principles screening across
the full chemical space is computationally infeasible, demanding an efficient, physically guided discovery strategy\cite{IE9,
	IE10}.

Machine learning has accelerated materials discovery in energy storage, catalysis, and structural
materials\cite{lu2023explainable, lopanitsyna2023modeling}, with explainable AI (XAI) methods increasingly revealing
structure-property relationships\cite{lu2025explainable}. However, ML application to quantum defect hosts faces unique
challenges: scarce experimental training data\cite{fiedler2022deep}, difficulty encoding multiscale defect physics in
compositional descriptors\cite{I16, I17, I18}, and the prevalence of opaque black-box models. Interpretability is particularly
critical for quantum hosts: when experimental synthesis is costly and coherence mechanisms remain incompletely understood, models
must reveal \textit{why} certain compositions succeed---extracting transferable physical principles (e.g., the role of dielectric
screening or nuclear-spin environments) that guide defect engineering and generalize beyond training data\cite{rudin_stop_2019,
	IE8, I18}.

Here, we present an interpretable machine learning framework using Rashomon set ensembles to screen host materials while
extracting physically grounded design principles\cite{rudin_stop_2019, molnar_2020}. We apply model-agnostic XAI tools, e.g.,
accumulated local effects (ALE), permutation feature importance (PFI), and individual conditional expectations (ICE) to understand
feature contributions across diverse classifiers\cite{molnar_2020}. Our approach combines structure-agnostic composition-only
screening with targeted first-principles validation. The framework rediscovers known hosts (diamond, \ce{SiC}, \ce{ZnO},
\ce{ZnS}) and identifies overlooked candidates (\ce{TiO2}, \ce{HfS2}, \ce{PbWO4}) across layered families suited to nanoscale
integration. By bridging data-driven prediction with physically interpretable insights, we establish a scalable pathway for
rational quantum materials discovery.

\section{Materials and Methods}

We integrate physically grounded machine learning with first-principles dielectric and defect calculations to identify
semiconductors that can host deep-level quantum-coherent spin defects. Our approach involves dataset construction using
coherence-informed filters, high-dimensional feature representation, explainable ensemble learning via Rashomon sets, and
DFT-based physical validation.

\subsection{Dataset Construction and Feature Representation}
We constructed a labeled dataset by intersecting the Materials Project\cite{MP1} and ICSD\cite{icsd1} databases. Only
thermodynamically stable phases with energy above the convex hull ($E_{\text{Hull}}$) less than 0.2 eV/atom were retained. Materials
containing noble gases were systematically removed. Positive class candidates were selected using coherence-relevant filters:
\begin{enumerate*}[label=(\alph*)]
	\item Bandgap $E_g > 0.5$ eV (to account for optical visibility and PBE underestimation);
	\item Compositions containing only elements with stable nuclear spin-zero isotopes;
	\item Absence of magnetic ordering; and
	\item Non-polar space groups to reduce inhomogeneous electric field noise\cite{IE8,M1,I17}.
\end{enumerate*}
Negative class samples were defined by contrasting conditions: magnetic, polar semiconductors ($ E_g > 0.1$ eV) containing at
least one element with non-zero nuclear spin. Polymorphs were merged to avoid redundancy. No structural overlap occurred between
the two classes. The final binary classification task exhibited class imbalance, addressed via stratified under-sampling during
training.

Materials retrieved from the database were reduced to unique compositions and redundant entries were removed. To
represent the materials in machine-learnable format, featurization was done using Matminer's tools \cite{matminer}.
Featurizers were chosen on a compositional-level\cite{MFE1,MFE2}, to represent a host material's electronic and chemical
environment. Any given composition is featurized and processed into 146 feature vectors using the descriptors,
\begin{enumerate*}[label=(\alph*)]
	\item Fractional elemental vectors;
	\item Valence electron counts and shell occupancy;
	\item Stoichiometric norms ($l_n$) capturing chemical heterogeneity\cite{ST};
	\item HOMO, LUMO Orbitals and Energies ($E_{HOMO}$ and $E_{LUMO}$), Atomic Orbitals, and corresponding Bandgap ($E_{AO}$)
	      derived from material composition\cite{AO}
	\item Estimate of the absolute Band Center ($E_C$) from composition \cite{BC}
\end{enumerate*}
to train the machine learning models.
Although stoichiometric features exhibited high inter-correlation (Pearson's $R > 0.9$), they were retained due to distinct
physical significance of individual features across models validated through Accumulated Local Effects (ALE) analysis\cite{Apley2020}.

\subsection{Ensemble Modeling and Rashomon Set - Analysis}
\begin{figure*}[htpb]
	\centering
	\includegraphics[width=\linewidth]{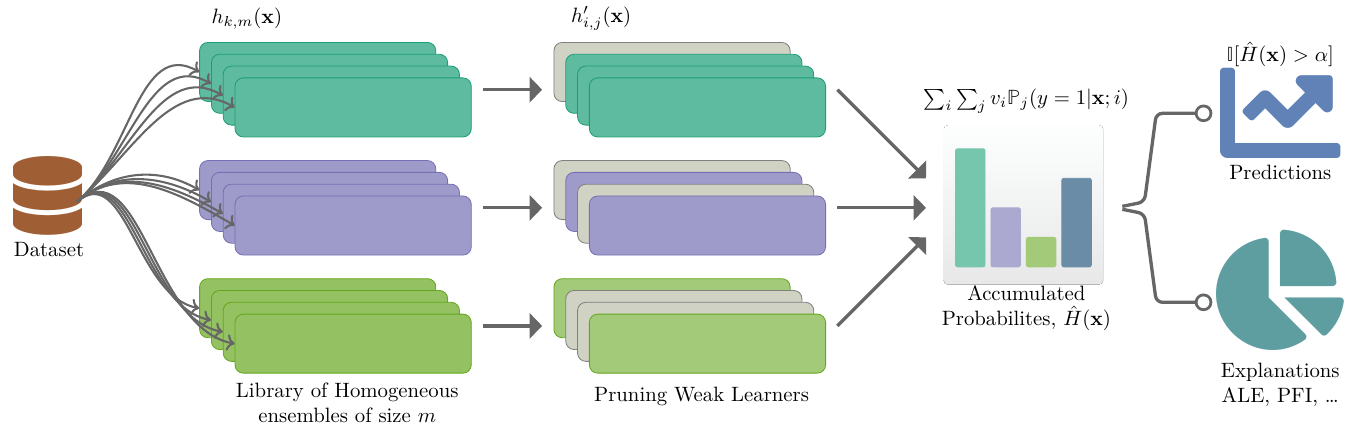}
	\caption{Schematic of ensemble construction and Rashomon filtering. Base learners trained by random undersampling of negative
		class data and the homogeneous ensembles are pruned to build a Rashomon set of predefined threshold $\epsilon$. Predictions are
		then made using the voting strategies discussed.}
	\label{fig:EC}
\end{figure*}

For predictive analysis, we trained diverse machine learning models including Stochastic Gradient Descent Classifier (SGC),
Random Forest (RFC), Support Vector Machine (SVC), K-Nearest Neighbors (KNN), Logistic Regression (LG), Bernoulli Naive Bayes
(NBC), and Gradient Boosting (GB). Optimal hyperparameters were selected via grid search using five-fold cross-validation. Given the
class imbalance, Matthews Correlation Coefficient (MCC) was used to reliably assess model generalization \cite{M4}. Despite
similar test performance, models exhibited distinct decision boundaries, forming Rashomon sets-subsets of models with comparable
accuracy but differing feature attributions \cite{rashomon, rashomon1}:

\begin{equation}\label{eq:2}
	\mathcal{R}(\epsilon,f^*,\mathcal{F}) = \{f\in\mathcal{F} \mid \text{MCC}(f)\ge \text{MCC}(f^*) - \epsilon \}
\end{equation}

where $f^*$ is the best performing model in the set and $\epsilon$ a tolerance threshold.

Two principal Rashomon sets emerged: $\mathcal{F}_1$ centered on SVC (including RFC, GB) and $\mathcal{F}_2$ around SGC
(including KNN, LG). We found that the models learned distinct and sometimes contradictory patterns from the same features.
Thus, the predictions of an individual model on Out of Distribution (OoD) data cannot be trusted. To enhance the
generalizability, we constructed a heterogeneous ensemble by pooling base learners from these sets \cite{MFE4}, as illustrated
in \figref{fig:EC}.

Obtaining meaningful interpretations from	Rashomon sets remains challenging especially in high-dimensional correlated feature
spaces \cite{rashomon2, rashomon3}. We addressed this by applying explainable AI (XAI) tools primarily relying on Permutation
Feature Importance (PFI) to identify influential features and Accumulated Local Effects (ALE) to disentangle their effects. The
learned patterns of individual models are then contrasted with ensemble-level	trends (see Section \ref{sec:xai}).

Training used a 7:2:1 split in an $\mathcal{M}\times\mathcal{N}$ grid, containing $\mathcal{M}$ types of homogeneous
ensembles each of size $\mathcal{N}$. Within each homogeneous ensemble, models were trained using stratified under-sampling.
The randomness generated by under-sampling the negative-class was found to provide sufficient diversity to the model
constituents. Predictions are made following,
\begin{enumerate*}[label=(\alph*)]
	\item \textbf{Mean Voting:} Average of predicted probabilities.
	\item \textbf{Constrained Voting:} A material is labeled positive only if all base ensembles (homogeneous ensembles) predict
	      $p > 0.5$.
\end{enumerate*}
This approach ensures robust, interpretable predictions by synthesizing diverse classifiers with complementary strengths.

\subsection{Unboxing the Black-box models}
\label{sec:xai}

To ensure reliable predictions, the decision functions of the base learners and the ensemble model were examined using
model-agnostic Explainable AI methods. Here, we discuss a brief overview of the XAI methods used. The Permutation Feature
Importance (PFI) calculates the change in predictive capability of a model when individual feature's contributions are
effectively hidden. There are a handful of implementations\cite{fisher_2019, rashomon2}, we used the simplest
one followed in sklearn's\cite{sklearn} inspection module. For a model, $\hat f:\mathcal{X}\longrightarrow\mathcal{Y}$, with
feature space $\mathcal{X} = \{\textbf x \mid \textbf x\in\mathbb{R}^n\}$, the Permutation Feature Importance is calculated by
randomly permuting the values of the feature of interest, $\textbf x_i$, in effect, replacing $\textbf x_i$ by a distribution that
is independent of $\textbf x_i$ but identical. The difference in scores of the model on the actual data and the permuted data is
the model reliance of the feature $mr(\textbf x_i)$ or the Permutation Feature Importance.
\begin{equation}\label{eq:3}
	mr(\textbf x_i) = S(\hat f(\mathcal{X})) - S(\hat f(\mathcal{X}'))
\end{equation}
The scoring function $S$ is chosen to be the Matthews Correlation coefficient\cite{chicco2023matthews}. PFI provides insights
about the feature's importance, but it doesn't explain the nature of the relationship. Further, if the feature space has
correlated vectors, the effect of permutation wouldn't have the desired effect since correlated features offer covariant
relationship to the model.

The Partial Dependence provides insights into the behavior of a set of features by considering a joint distribution of the
dataset with the complement subset of the feature space taking on average values\cite{friedman2001greedy, molnar_2020}.
The Partial Dependence of a subset of features $\textbf X_i$, is given by,
\begin{equation}\label{eq:4}
	\hat f_{\textbf X_j}({\textbf X_i}) = \mathbb{E}_{\textbf X_j}\left[ \hat f(\textbf X_i, \textbf X_j)\right] =\int\hat
	f(\textbf X_i, \textbf X_j)d\mathbb{P}(\textbf X_j)
\end{equation}
Where $\textbf X_j$ is the complement of the feature space such that $\mathcal{\textbf X}=\textbf X_i\cup \textbf X_j$. For
practical applications, the integral is approximated by considering the summation over a desired range,
$$ \int\hat	f(\textbf X_i, \textbf X_j)d\mathbb{P}(\textbf X_j) = \frac{1}{n}\sum_{k=1}^n 	f(\textbf X_i, \textbf X_j^{(k)}) $$
The individual $k$ corresponds to an instance in the complement dataset $\textbf X_j$. ICE curves\cite{goldstein2015peeking}
utilize these values to show the effect of the set $\textbf X_i$ per instance. In the ICE plots in
\figref{fig:ICE_S}, the conditional expectations $f(\textbf X_i, \textbf X_j^{(k)})$ are plotted for 200 different values
of $k$ spread equidistant throughout the feature's range corresponding to 200 curves. The average indicated by the purple line,
is the approximated Partial Dependence Curve. Partial Dependence is however affected by correlations in the feature space,
producing misleading data points in the feature set.

The Accumulated Local Effects\cite{apley2020visualizing} (ALE) unlike traditional XAI methods involving global conditional expectations
such as PDP and SHAP(SHapley Additive exPlanations)\cite{lundberg2017unified}, is robust to correlations, and
shows the average effect of a feature accumulated in small intervals over its entire range. The ALE of a feature $X_i$ with
range $(a,b)$ is computed by averaging the local effects of the feature, $\hat f_{ij}(X_i, X_{\backslash j})=\frac{\delta
		\hat f(\textbf X_i, X_j)}{\delta
		\textbf X_i}$ in grids of small intervals, across the range, $(a,b)$. We chose Alibi's implementation \cite{alibi}
that uses finite differences to compute and accumulate the effects\cite[Chapter~20]{molnar_2020} as,

\begin{equation}\label{eq:5}
	\begin{aligned}
		\hat f_{j,\text{ALE}}(X_i)= & \sum_{k=1}^{k_j(x)}\frac{1}{n_j(k)}\sum_{i:X_j^{(i)}\in N_j(k)} \\ & \left[ \hat f_{kj}(Z_{k,j}, X_{\backslash j})- f_{k-1,j}(Z_{k-1,j}, X_{\backslash j})\right]
	\end{aligned}
\end{equation}
The model's local effects $\hat f$ on the distribution in the grid $Z$ are accumulated in multiple intervals of varying
population $N_j$ and then averaged over the whole range. Effects of correlations are transparent in the grids.

\subsection{First-Principles Validation and Dielectric Calculations}
To validate the physical viability of the predicted host materials, density functional theory (DFT) calculations were carried out using VASP\cite{Kresse1996,Kresse1999} with PAW potentials\cite{Blochl1994} and the PBE exchange-correlation functional\cite{Perdew1996}. Structures were relaxed until forces converged below 0.01 eV/\AA{} using plane-wave cutoffs of 520 eV and Monkhorst-Pack $k$-point grids (at least $5 \times 5 \times 5$).

Defect formation energies $E_f[D^q]$ were calculated for oxygen vacancies using $2 \times 2 \times 2$ supercells, corrected using
the Freysoldt-Neugebauer-Van de Walle (FNV) scheme\cite{Freysoldt2009}:

\begin{equation}\label{eq:6}
	\begin{aligned}
		E_f[D^q] = & E_\mathrm{tot}[D^q] - E_\mathrm{tot}[\text{bulk}] -          \\
		           & \sum_i n_i \mu_i + q(E_F + \varepsilon_v) + E_\mathrm{corr},
	\end{aligned}
\end{equation}

where $E_\mathrm{corr}$ includes image charge corrections\cite{Zhang1991,Lany2008}. Vacancy and interstitial defect sites were
identified using pymatgen\cite{MP1}.

Dielectric constants were obtained using density functional perturbation theory (DFPT)\cite{Baroni_1986,DFPT_2006}, with explicit separation of electronic ($\varepsilon_\infty$) and ionic ($\varepsilon_{\text{ion}}$) components. High total dielectric constants, particularly with dominant ionic response, are known to mitigate electric field noise and enhance spin coherence\cite{M1, I17}.

All results from DFT and DFPT calculations are used to cross-validate the ML-predicted candidate materials and assess their
suitability as defect-tolerant quantum hosts. Additional implementation details on data preprocessing, Ensemble pruning, feature
analysis are provided in the Supplementary Information.

\section{Results}

\subsection{Model Performance and Interpretability}
\begin{figure*}[ht]
	\begin{subfigure}{0.68\textwidth}
		\centering
		\begin{subfigure}{\textwidth}
			\includegraphics[width=\textwidth]{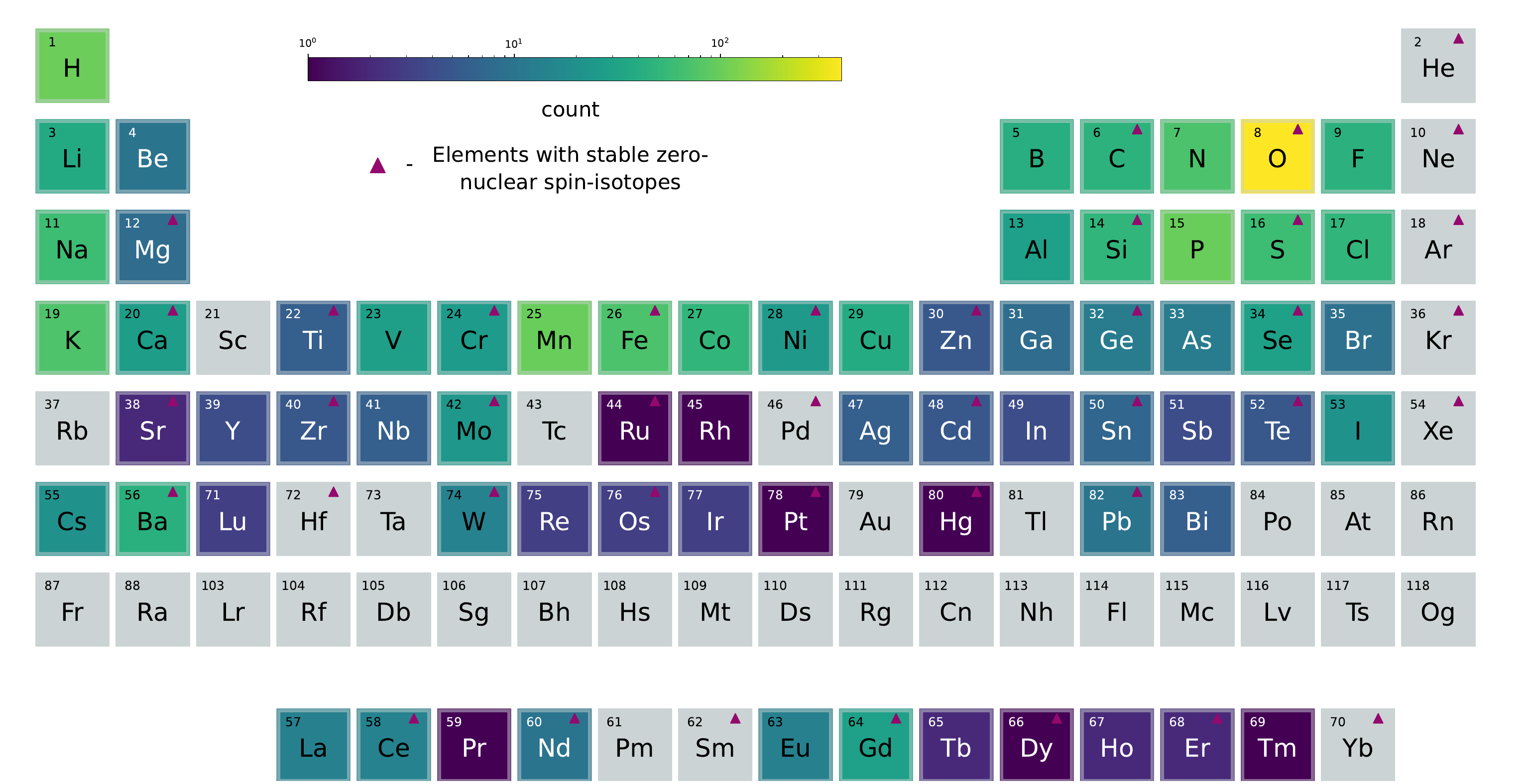}
			\caption{}
			\label{fig:data}
		\end{subfigure}
		\begin{subfigure}[t]{\textwidth}
			\centering
			\includegraphics[width=\textwidth]{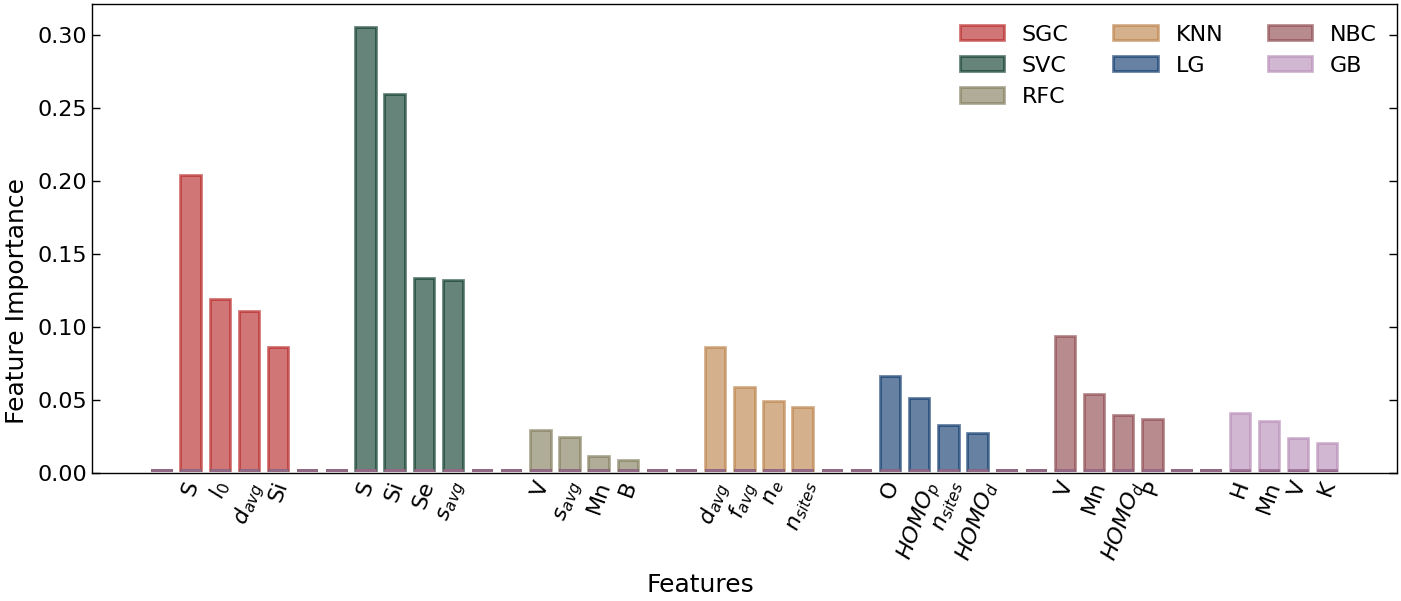}
			\caption{}
			\label{fig:pfi_plots}
		\end{subfigure}
	\end{subfigure}
	\begin{subfigure}{0.3\textwidth}
		\begin{subfigure}{\textwidth}
			\includegraphics[width=.9\textwidth]{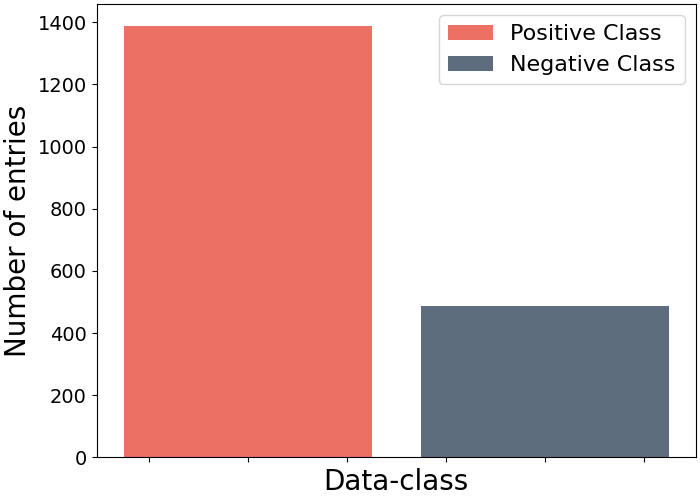}
			\caption{}
			\label{fig:dist_elem}
		\end{subfigure}
		\begin{subfigure}{\textwidth}
			\includegraphics[width=.9\textwidth]{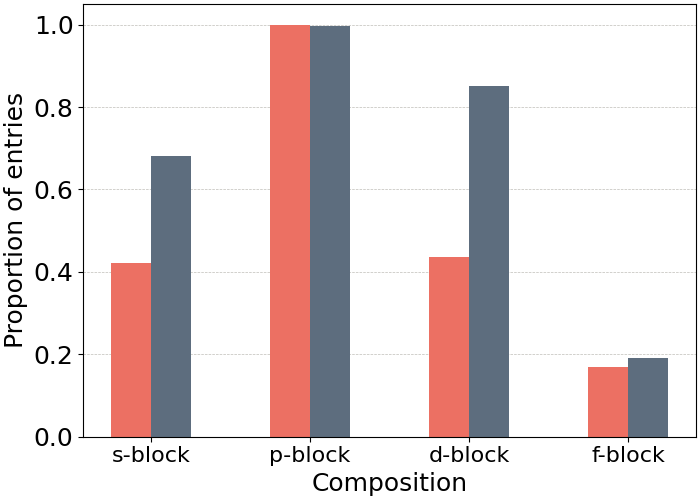}
			\caption{}
			\label{fig:dist_elec}
		\end{subfigure}
		\begin{subfigure}{\textwidth}
			\includegraphics[width=.9\textwidth]{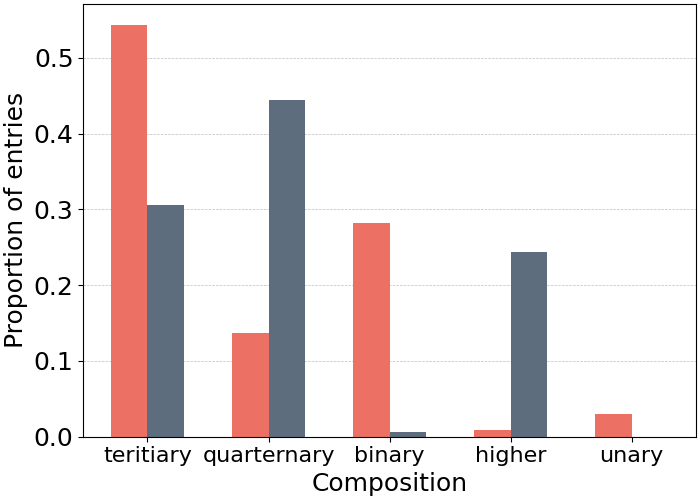}
			\caption{}
			\label{fig:dist_comp}
		\end{subfigure}
	\end{subfigure}
	\caption{(a) Element-wise distribution of negative class data. By restricting the positive class to materials
		that exclusively contain elements with stable zero nuclear spin isotopes, the distribution of elements is severely
		restricted.  But an identical distribution of elements across the binary classes prevents any bias towards specific element
		species.
		(b) Permutation Feature Importance (PFI) highlighting the most influential features for
		different models.  Matthews correlation coefficient (MCC) is used as the metric to evaluate the feature importance.
		Individual models rank different features to be important for predictive analysis.  The set of features considered for
		exploration using XAI methods were derived partly from PFI-ranked features and partly using ALE (Accumulated Local
		Effect)-analysis.
		(c) Class imbalance in the retrieved data.
		(d) Distribution of
		data as per the cumulative valence electrons in each instance. Skewness in the distribution would manifest in the feature space
		as spurious patterns and may override a model's generalizable decision boundary.
		(e) element species-wise
		distribution of compositions, giving an overview of $l_0$ distribution in the feature space.}
	\label{fig:distribution}
\end{figure*}

The dataset was curated from the Materials Project using coherence-relevant filters such as wide bandgap and
zero nuclear spin-bath. This filtering inherently biases the dataset toward the positive class, creating class imbalance
and potentially specious feature-space separations. To assess and mitigate sampling biases, we examined the feature-space
distribution in \figref{fig:distribution}.

Restricting the positive class to nuclear-spin-zero elements severely limits elemental diversity, the sparse nature of
element fraction vectors could result in the development of artificial decision boundaries if a non-intersecting subset of
elements make up the binary classes. As shown in \figref{fig:data}, the negative class spans the periodic table, forming an
intersecting set with the
positive class. Individual models rank different features as important, underscoring the Rashomon nature: complex models
(Random Forest, Gradient Boosting) over-rely on element fractions and generalize poorly to under-represented compositions,
while simpler models (KNN) learn more robust attributes as shown in \figref{fig:pfi_plots}. Along with element fraction
vectors, valence electron counts and $s$,	$d$, $f$ orbital occupancy emerge as key classifiers.

Despite reasonable balance in valence-electron distributions (\figref{fig:dist_elec}), compositional weights are skewed
(\figref{fig:dist_comp}), with the negative class lacking unary and binary compounds. While this reflects well-known trends
in quantum compatible materials (experimentally verified hosts are unary or binary; \tabref{tab:dft}), to look past simple
heuristics, stoichiometric features should capture more latent information. ALE analysis revealed that EC
learns non-linear relationships from the stoichiometric features ($l_n$-norms). To address the class imbalance, we
employ a heterogeneous ensemble that strategically undersamples the positive class across base learners while retaining full
training data, effectively mitigating emergent biases.

\begin{figure*}[!ht]
	\centering
	\begin{subfigure}{.45\linewidth}
		\includegraphics[width = .95\textwidth]{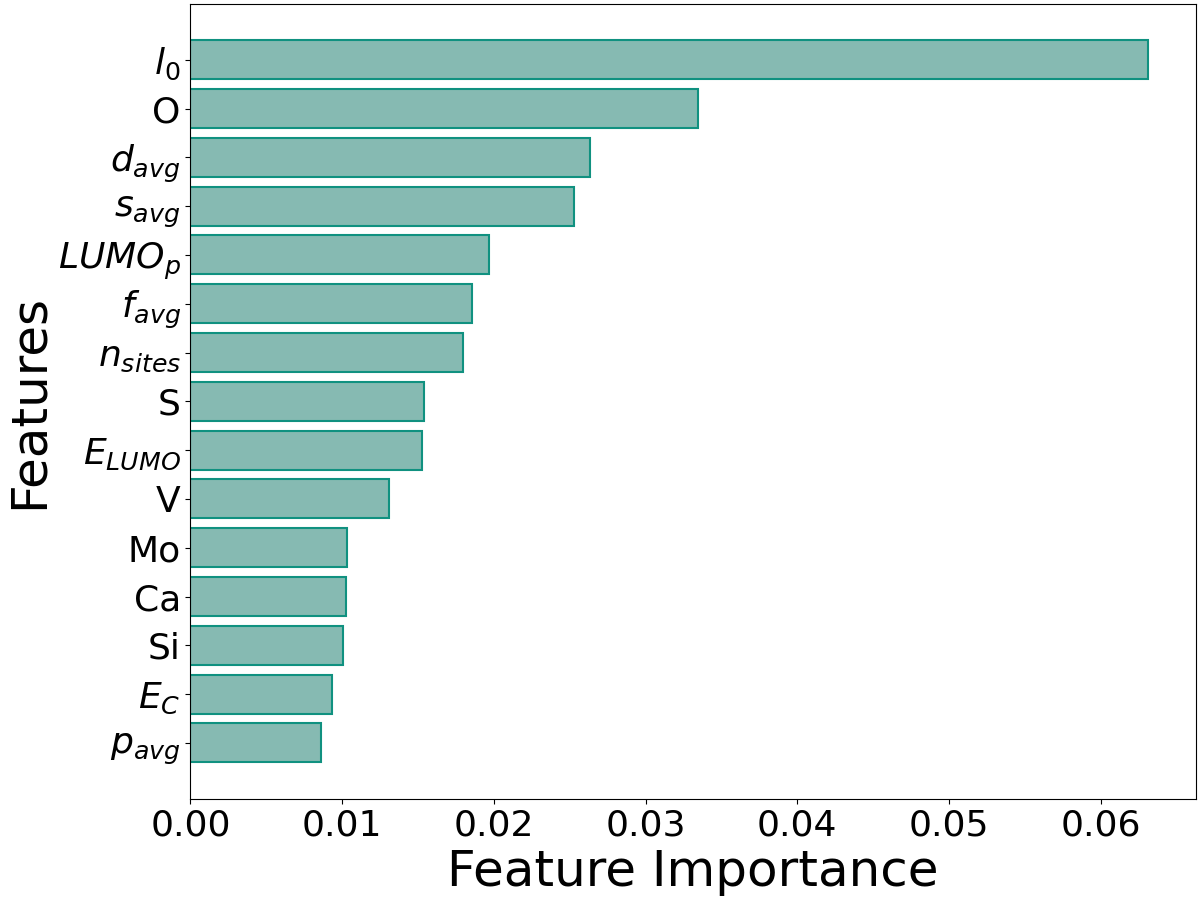}
		\caption{}
		\label{fig:EC_pfi}
	\end{subfigure}
	\begin{subfigure}{.45\linewidth}
		\includegraphics[width = .95\textwidth]{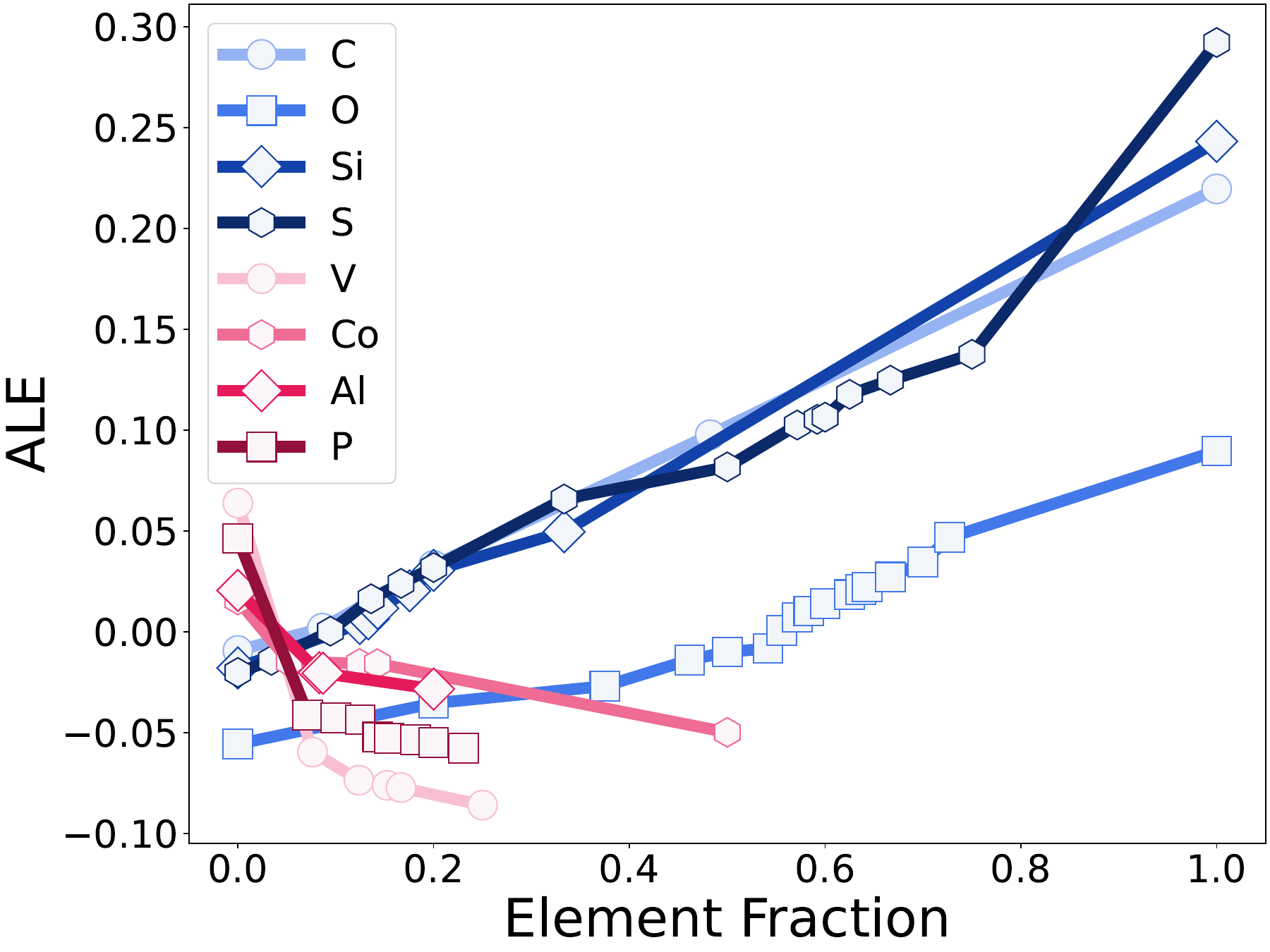}
		\caption{}
		\label{fig:EF}
	\end{subfigure}
	\begin{subfigure}{.45\linewidth}
		\includegraphics[width = .95\textwidth]{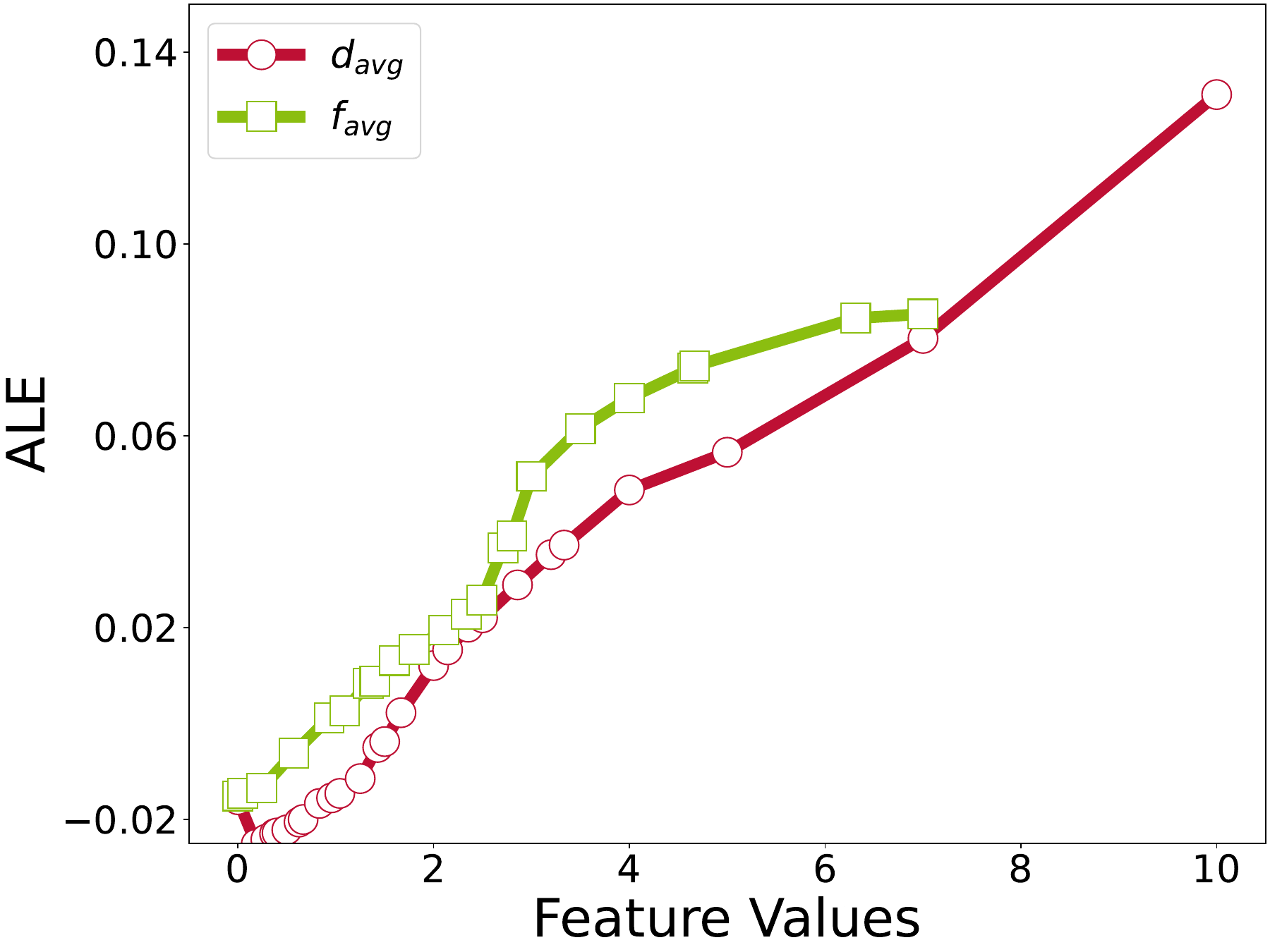}
		\caption{}
		\label{fig:EF_df}
	\end{subfigure}
	\begin{subfigure}{.45\linewidth}
		\centering
		\includegraphics[width = .95\textwidth]{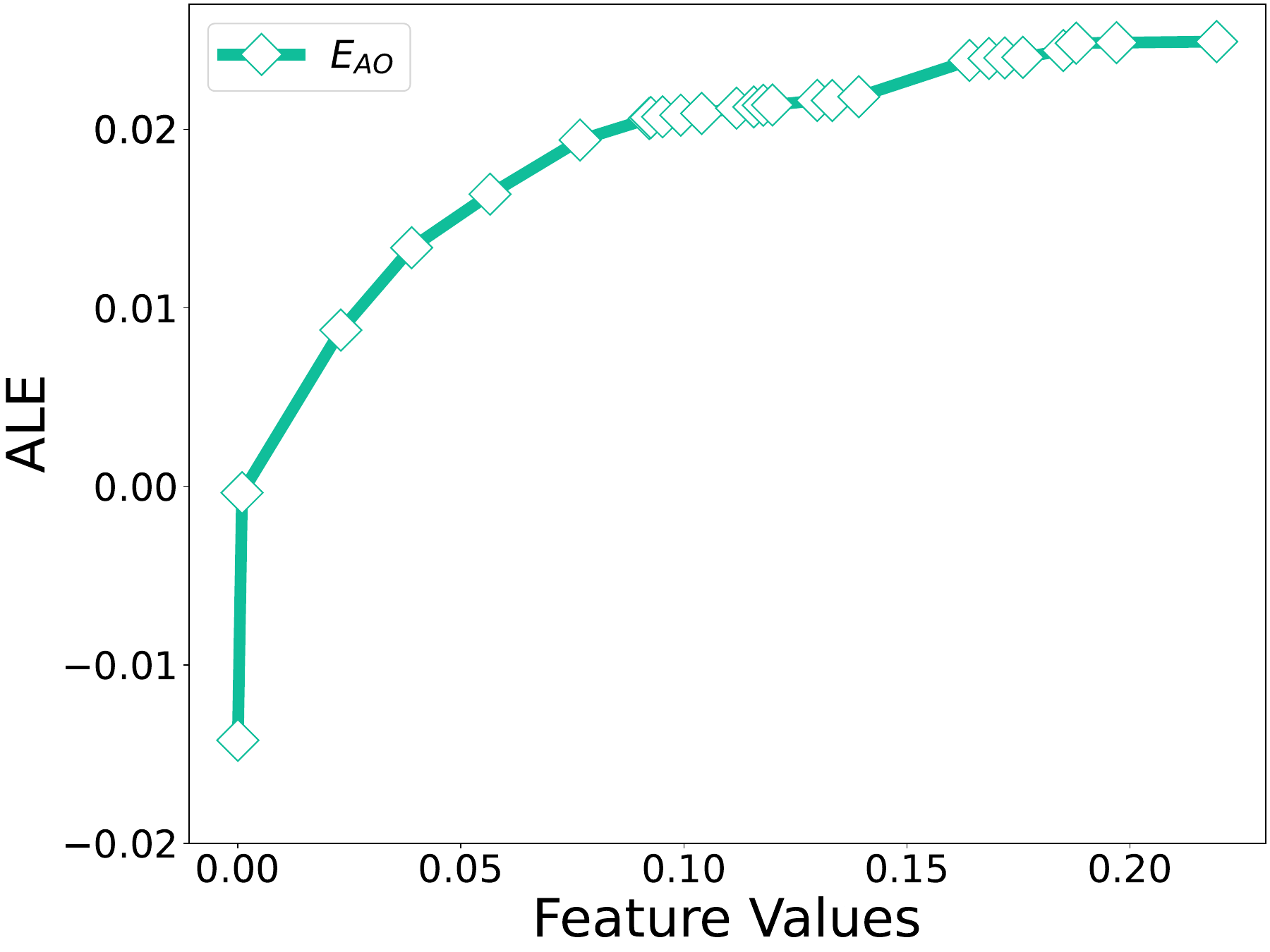}
		\caption{}
		\label{fig:EF_E_AO}
	\end{subfigure}
	\caption{
		(\subref{fig:EC_pfi}) Permutation Feature Importance (PFI) listing features that EC relies on significantly to classify a given composition.
		(\subref{fig:EF}) The effect of element fractions on positive classification probability. EC assigns higher likelihood to compositions containing \ce{C}, \ce{S}, \ce{Si}, and \ce{O}, and de-emphasises compositions containing \ce{Co}, \ce{V}, \ce{P}, \ce{Al}, and \ce{Mn}.
		(\subref{fig:EF_df}) The effect of average number of \textit{d} ($d_{avg}$) and \textit{f} ($f_{avg}$) valence electrons in a composition. On average, compositions with fully filled \textit{d} and \textit{f} configurations have a higher likelihood of being quantum compatible.
		(\subref{fig:EF_E_AO}) The effect of bandgap estimated from atomic orbital energies ($E_{AO}$) on positive classification. EC prefers wide bandgap compositions.
		In (\subref{fig:EF})--(\subref{fig:EF_E_AO}), ALE on positive predicted probability that a given instance is a compatible host is plotted on the vertical axis with the feature's range on the horizontal axis.}
	\label{fig:EC_features}
\end{figure*}

The ensemble was trained on a $7 \times 400$ model grid spanning seven classifier types, then pruned and combined using a mean-voting scheme
to compute predictions. The resulting Ensemble Classifier (EC) achieved an MCC of $\approx$0.99 with near-perfect precision,
recall, and F1 score on the held-out test set, outperforming all individual base learners and demonstrating the effectiveness of
the heterogeneous ensemble strategy (\tabref{tab:learning_curve}). However, these metrics alone do not guarantee
generalizability: accumulated local effects (ALE) analysis shows that models within the same Rashomon set can learn contrasting
relationships from identical features.  Instead of selecting a single best model, the heterogeneous ensemble synthesizes these
differing hypotheses via constrained voting, yielding predictions only when multiple independent reasoning pathways agree. Even
under average voting, EC recognizes features important to both Rashomon sets (\figref{fig:EC_pfi}); the effects of
these features were subsequently examined using ALE analysis.

\begin{table}
	\centering
	\resizebox{.95\linewidth}{!}{\begin{tabular}{lcrrrrrrr}
			\toprule
			Metrics         & SGC   & SVC   & RFC   & KNN   & LG    & NBC   & GB    & EC$^*$ \\
			\midrule
			Train Precision & 0.988 & 1.000 & 1.000 & 1.000 & 0.957 & 0.965 & 0.996 & 1.000  \\
			Test Precision  & 0.963 & 0.983 & 0.961 & 0.909 & 0.903 & 0.944 & 0.934 & 0.992  \\
			Train Recall    & 0.911 & 1.000 & 1.000 & 1.000 & 0.983 & 0.986 & 1.000 & 1.000  \\
			Test Recall     & 0.913 & 0.994 & 1.000 & 0.983 & 0.977 & 0.978 & 0.994 & 1.000  \\
			Train F1 Score  & 0.948 & 1.000 & 1.000 & 1.000 & 0.971 & 0.976 & 0.998 & 1.000  \\
			Test F1 Score   & 0.937 & 0.988 & 0.980 & 0.944 & 0.939 & 0.960 & 0.963 & 0.996  \\
			Train MCC       & 0.864 & 1.000 & 1.000 & 1.000 & 0.913 & 0.928 & 0.928 & 1.000  \\
			Test MCC        & 0.836 & 0.968 & 0.944 & 0.839 & 0.823 & 0.887 & 0.888 & 0.985  \\
			\bottomrule
		\end{tabular}}
	\caption{Performance matrix of the base models and the ensemble classifier (EC$^*$) on training and test datasets. EC has an
		overall better MCC score in the test set while other metrics are identically distributed in models belonging to the
		$\mathcal{F}_1$ set. MCC shows a more reliable estimate of how well a model would generalize on unseen data.}
	\label{tab:learning_curve}
\end{table}

\begin{figure*}[ht]
	\centering
	\includegraphics[width=\linewidth]{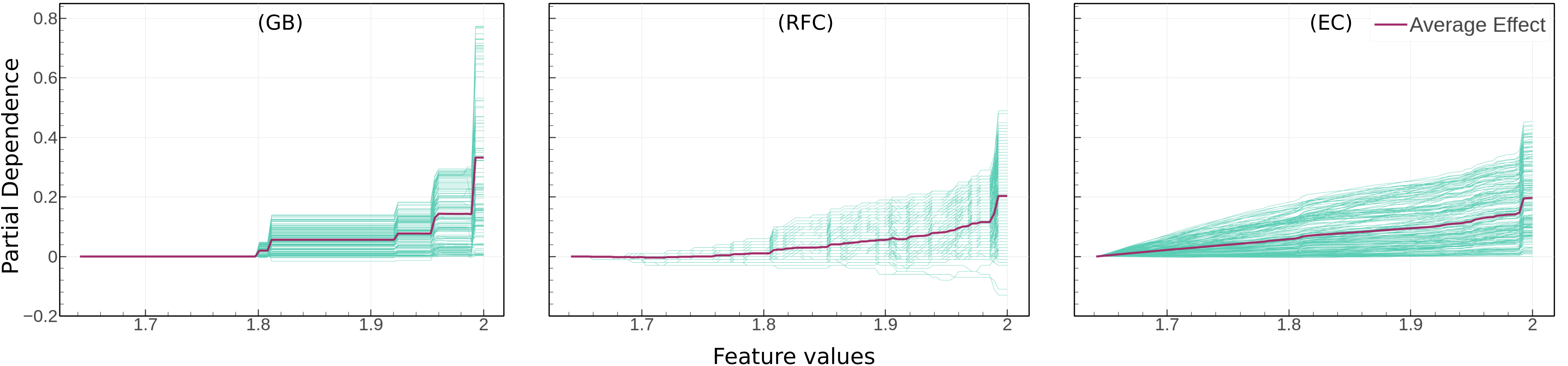}
	\caption{\centering ICE Plots of the feature $s_{avg}$, the average number of \textit{s} valence electrons in a composition. The models
		agree that a filled \textit{s}-orbital in the valence shells is beneficial for a host material to be quantum compatible.}
	\label{fig:ICE_S}
\end{figure*}

\begin{figure*}[htpb]
	\centering
	\includegraphics[width=\textwidth]{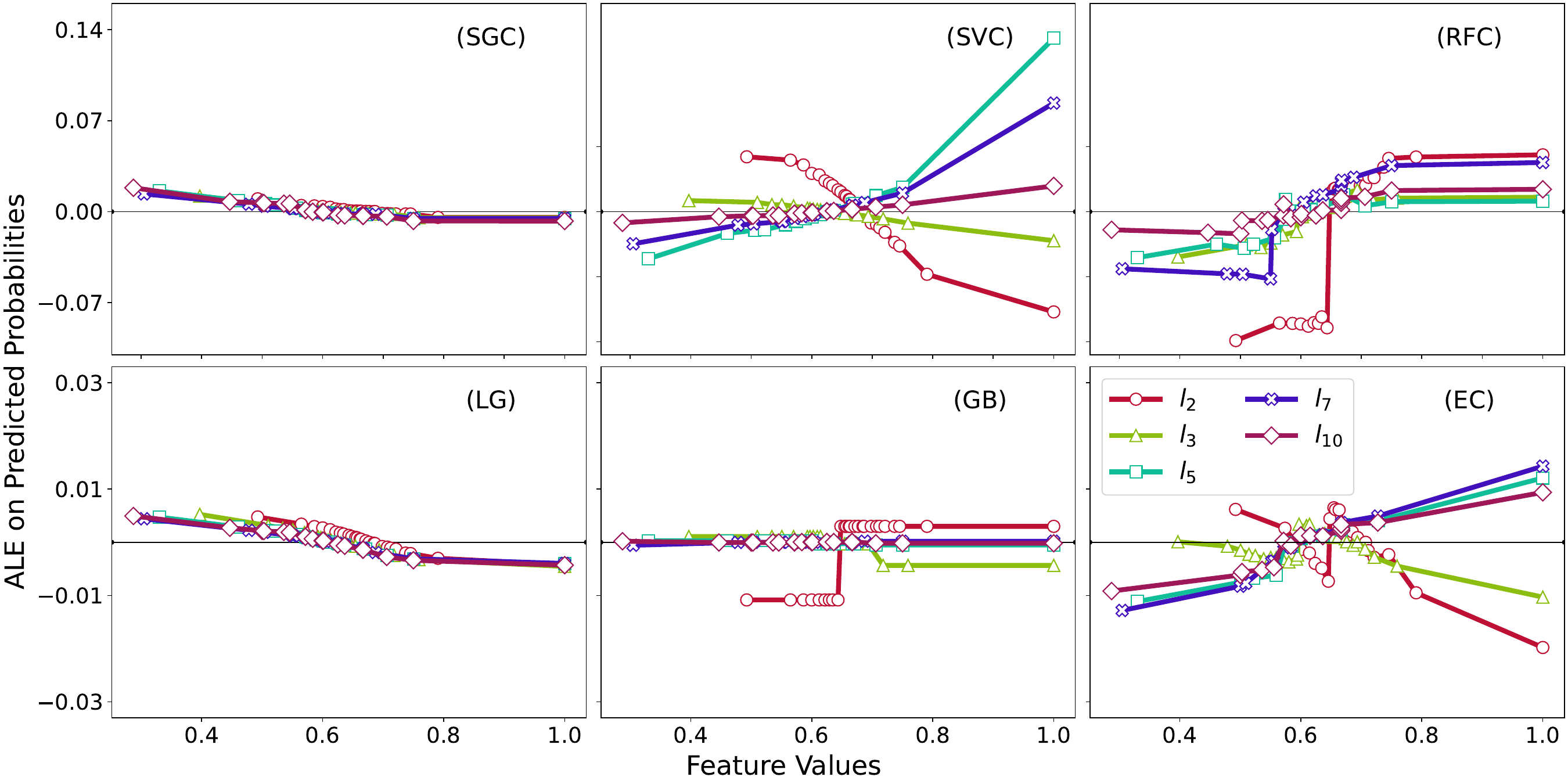}
	\caption{\centering Accumulated local effects (ALE) plots highlighting the global effects of highly correlated stoichiometric features
		on predicted probabilities. $l_n$ quantifies the chemical heterogeneity of a composition, with higher values of $n$ giving
		more importance to the most prominent element. While the linear base line model learned the same relationships from these
		features, despite being correlated, the features introduce diverse and sometimes contradictory patterns to the models. Even
		models from the same Rashomon set learned disparate relationships over a wide effect range. This will lead to significantly
		different predictions for the same instance. The ensemble model assigns higher probability to materials with low chemical
		heterogeneity indicated by the overall increasing trend in higher order norms ($l_7$ and $l_{10}$) and prefers simpler
		compositions indicated by the declining trend of lower order norms.
	}
	\label{fig:ale_plots}
\end{figure*}

To explain the decision functions of the models, we used the model-agnostic XAI methods described in Section~\ref{sec:xai} to
examine the feature effects. To only include meaningful effects, we consider the set of features recognized to be important by the
models by analyzing their PFI values. From this subset, we examine
patterns on which all models broadly agree. Although feature rankings differ, the base estimators concur that, on average,
simpler compositions are preferred (\figref{fig:ale_plots}), a completely filled \textit{s}-orbital markedly improves quantum
compatibility (\figref{fig:ICE_S}), and the presence of elements such as carbon, sulfur, and oxygen increases the likelihood
that a material belongs to the positive class. Most models agree that carbon and silicon generally contribute positively to
quantum compatibility, albeit with varying effect sizes, but contradict one another on the role of oxygen and sulfur. Confident
individual models like RFC, SVC, and GB fail to discern known patterns, such as the beneficial influence of nuclear-spin-less
binary oxides on coherence times \cite{mgo}. In contrast, the Ensemble model successfully identifies these trends
(\figref{fig:EF}), particularly unary compositions of silicon and sulfur which, like diamond, are predicted to host
quantum-compatible deep-level defects \cite{kanai}. EC's hypotheses on element-fraction vectors are consistent with ab-initio
studies: Kanai \textit{et al.}\cite{kanai} report an abundance of sulfates, sulfides, and oxides among potential hosts from
more than 12,000 materials, and EC assigns relatively higher probability to sulfur-related compounds than to carbides and
silicates (\figref{fig:EF}). This feature exploration narrows the chemical search space for high-throughput defect
calculations; binary and ternary compositions containing \ce{S, C, Si}, and \ce{O} are most likely to form homogeneous
nuclear-spin baths supporting longer coherence times.

We extend this analysis to other features highlighted by EC's PFI. EC de-emphasizes compositions containing \ce{Co, V, P, Al},
and \ce{Mn} (\figref{fig:EF}). While intrinsic magnetic moments in \ce{Mn} and \ce{Co} are known to degrade coherence,
similar behavior observed for \ce{Al} and \ce{P}-based compositions could point to a latent trend that has yet to be identified.
ALE analysis of the average number of \textit{d} and
\textit{f} valence electrons (\figref{fig:EF_df}) shows that EC prefers compositions with completely filled \textit{d} and
\textit{f} shells. Combined with the oxygen and sulfur fraction effects in \figref{fig:EF}, this points to materials such as
\ce{PbS}, \ce{WO3}, and \ce{Pb2WO5} as promising hosts. While \ce{WO3} is expected to exhibit $T_2$ times exceeding diamond
\cite{kanai}, lead sulfides and oxides remain largely unexplored. EC further identifies a positive correlation between the
bandgap estimated from atomic orbital energies, $E_{AO}$, and positive classification, though its effective range is narrower
than that of valence-electron attributes (\figref{fig:EF_E_AO}). Correlated features such as $E_{AO}$ and $f_{avg}$ are often
downplayed by PFI, PFI being a popular method to assess the feature importance tends to suppress correlated descriptors, yet ALE
reveals that they exert distinct, physically meaningful effects on the decision function.

Finally, ALE analysis of highly correlated stoichiometric norms (\figref{fig:ale_plots}) uncovers distinct, physically
interpretable relationships across base learners. Models within the same Rashomon set learn contrasting patterns over wide
effect ranges, leading to divergent predictions for the similar instances, highlighting the need to examine global feature
effects for trustworthy inference. The ensemble classifier EC, by aggregating these hypotheses, identifies consistent trends:
simpler compositions with reduced elemental diversity and lower chemical heterogeneity (higher-order $l_n$ norms) have a higher
probability of being classified as quantum-compatible hosts. These hypotheses align with established coherence
criteria that the chemical homogeneity minimizes nuclear-spin interactions.

\subsection{Screening of Quantum-Compatible Host Materials}
We applied the ensemble classifier to $\sim$45,000 thermodynamically stable Materials Project compounds ($E_{\text{Hull}}<0.2$ eV/atom) using constrained voting (Section II.B), identifying 122 high-confidence candidates at $>0.95$ threshold (\figref{fig:chart}, \tabref{tab:predictions}).

\subsubsection{Threshold Selection and Model Comparison}

\begin{figure*}[htpb]
	\begin{subfigure}[c]{.42\textwidth}
		\begin{center}
			\includegraphics[width=\textwidth]{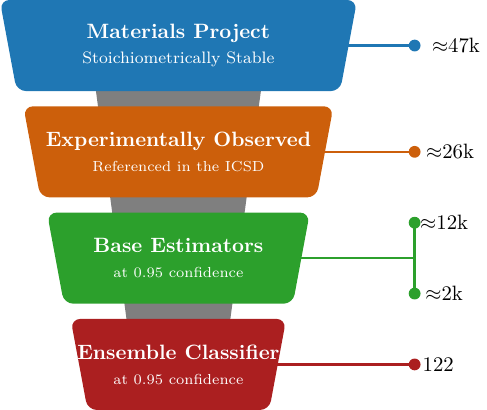}
		\end{center}
		\caption{}
		\label{fig:chart}
	\end{subfigure}
	\begin{subfigure}[c]{.55\textwidth}
		\begin{center}
			\includegraphics[width=.9\textwidth]{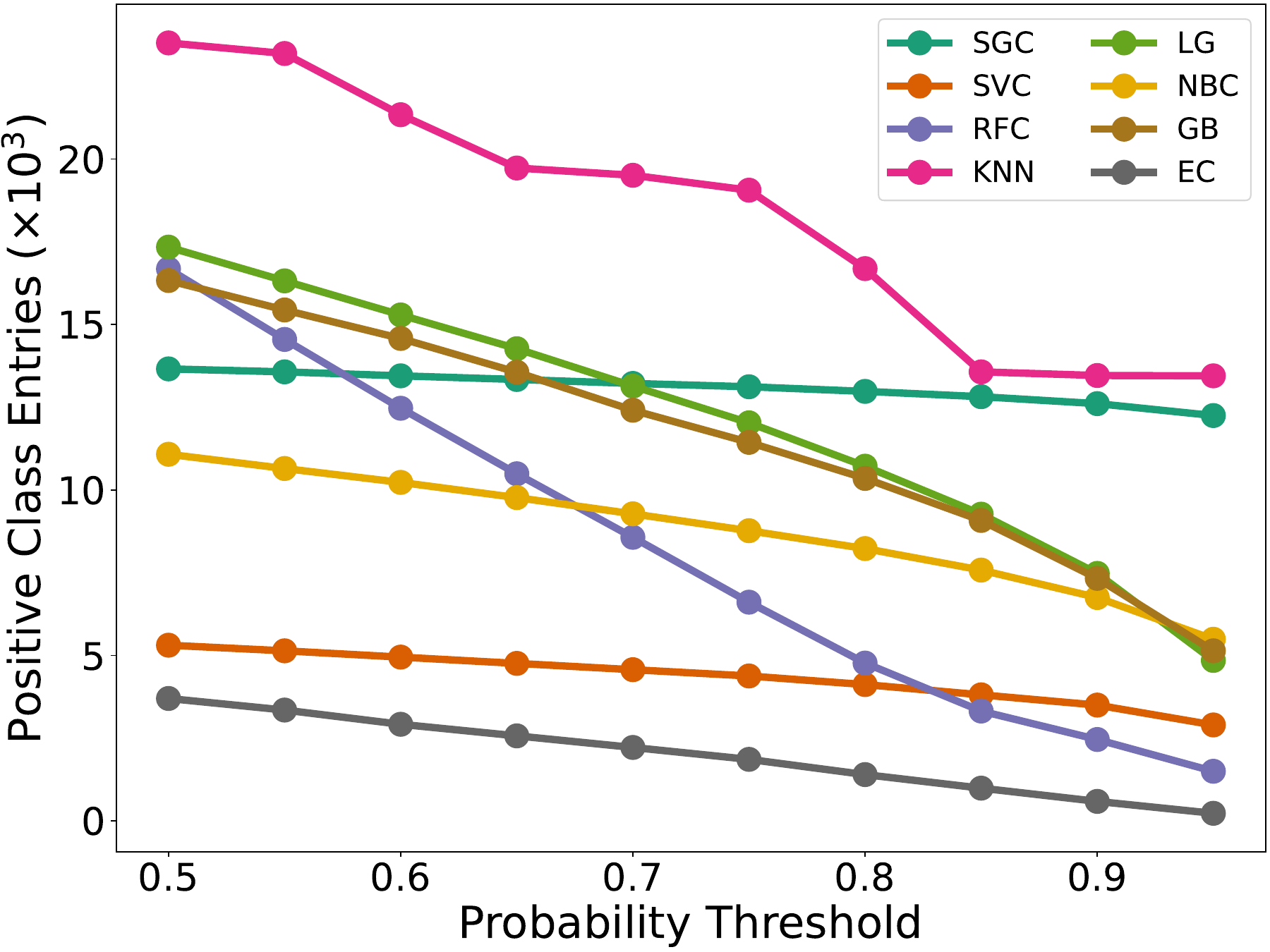}
		\end{center}
		\caption{}
		\label{fig:num_entries}
	\end{subfigure}
	\caption{(\subref{fig:chart}) Distribution of the 122 high-confidence ($>0.95$) predictions by chemical family, showing the
		dominance of oxides and chalcogenides alongside smaller contributions from carbides, tellurides, and ternary compounds.
		(\subref{fig:num_entries}) Number of positive predictions from the Materials Project database as a function of ensemble
		confidence threshold for the ensemble classifier (EC) and individual base learners (NBC, RFC).  EC produces a sharper transition
		than any single model: at 0.95, it yields 122 candidates---intermediate between the conservative RFC and the permissive NBC
		---while retaining most experimentally verified hosts.}
	\label{fig:flow}
\end{figure*}

The confidence threshold governing candidate selection represents a critical methodological choice that directly determines the
precision--recall balance of the screening.  \figref{fig:num_entries} compares the number of positive predictions across
confidence thresholds for the ensemble classifier (EC) and its constituent base learners.  At a threshold of 0.90, EC identifies
$\sim$500 candidates; this set, however, includes narrow-bandgap phases (e.g., several tellurides and germanides with $E_g <
	0.5$\,eV) and compositions containing elements with non-negligible paramagnetic susceptibility, both of which are unlikely to
sustain long spin coherence.  Raising the threshold to 0.95 reduces the candidate pool to 122 while retaining most
experimentally verified quantum host (see below), effectively filtering out materials that satisfy the compositional criteria
only marginally.  A still more stringent cutoff of 0.99 would discard $\sim$40\% of these 122 candidates, including several oxide
and chalcogenide families with strong theoretical support\cite{kanai, M1MgO}, without a commensurate gain in precision.  The 0.95
threshold thus represents the point at which recall of known hosts is complete and the marginal materials entering the candidate
set are compositionally ambiguous rather than clearly incompatible.

The ensemble sharpens predictions relative to individual models: at 0.95, RFC yields $>$1000 candidates and NBC yields
$\sim$5000, reflecting their complementary biases---RFC is conservative, penalizing compositions that deviate from the
training-set norm, whereas NBC over-classifies borderline materials since it is insensitive to element-fraction features.  The
heterogeneous ensemble, by requiring agreement across these divergent decision boundaries, suppresses both false positives from
NBC and false negatives from RFC, yielding a candidate set that is simultaneously more selective and more complete than any
single constituent model.

\subsubsection{Validation Against Known Hosts}

The primary test of a screening framework is whether it recovers materials already known to work.  Among the 122 candidates at
the 0.95 threshold, the ensemble correctly identifies all major experimentally verified quantum defect hosts: diamond (\ce{C},
confidence 0.96)\cite{I15NVCenter}, silicon carbide (\ce{SiC}, 0.97)\cite{qian2025first}, zinc oxide (\ce{ZnO},
0.98)\cite{M1ZnO}, and zinc sulfide (\ce{ZnS}, 0.98)\cite{hoang2022rare}.  The layered transition metal dichalcogenides \ce{WS2}
(< 0.9)\cite{I10} and \ce{MoS2} (< 0.8)\cite{M1TMD} appear at lower threshold values. Compositionally, these hosts contain heavy
transition metals (\ce{W}, \ce{Mo}) whose partially filled $d$-shells place them near the decision boundary between the two
Rashomon sets, consistent with the ALE analysis showing that EC assigns reduced probability to compositions with intermediate
$d$-electron counts (\figref{fig:EF_df}).  That these hosts are recovered at all---albeit at reduced confidence---validates the
model's ability to generalize to chemistries where coherence arises from structural rather than purely compositional factors.

Beyond the hosts verified by direct experiment, the 122-candidate set also recovers materials identified as promising by
independent computational studies: \ce{MgO}, \ce{CaO}, \ce{WO3}, and \ce{CaS} from the large-scale DFT screening of Kanai
\textit{et al.}\cite{kanai}, and \ce{MgO} from the defect-level calculations of Davidsson \textit{et al.}\cite{M1MgO}, all at
confidence $\geq 0.95$.  The recent prediction that \ce{SiO2} hosts several promising defect configurations\cite{lang2025solid}
is likewise recovered, providing an independent cross-check between our composition-only approach and structure-resolved
screening.  In total, the model reproduces candidates from at least three methodologically distinct studies without having been
trained on their outputs, suggesting that the learned decision boundary captures transferable physical criteria rather than
dataset-specific correlations.  Recovering 100\% of known hosts at high confidence is non-trivial given the compositional
diversity of these materials---spanning unary (C), binary ionic (MgO, CaO, CaS), binary covalent (SiC, ZnO, ZnS), and layered van
der Waals (WS$_2$, MoS$_2$) bonding environments---and indicates that the ensemble has learned to weight the underlying physical
drivers (nuclear-spin-bath purity, dielectric screening capacity, electronic structure compatibility) rather than superficial
chemical similarities.

\subsubsection{Novel Predictions by Chemical Family}
Our predictions can be categorized in four categories:
\begin{enumerate*}[label=(\Roman*)]
	\item \textit{Layered chalcogenides:} \ce{HfS2}, \ce{ZrS2}, \ce{SnS2}, \ce{TiS2}-heavier analogs of studied \ce{WS2}/\ce{MoS2} with favorable nuclear-spin properties for vdW heterostructures.
	\item \textit{High-$\varepsilon$ oxides:} \ce{TiO2}, \ce{BaO}, \ce{SrO}, tungstates (\ce{PbWO4}, \ce{CdWO4}), molybdates-previously unexplored for spin defects.
	\item \textit{Alkaline earth tellurides:} \ce{CaTe}, \ce{SrTe}, \ce{BaTe}, \ce{PbS}/\ce{PbSe}-predicted favorable despite narrow-gap phases (strain/alloying may stabilize wide-gap polymorphs).
	\item \textit{Ternaries:} Zirconates (\ce{SrZrO3}, \ce{CaZrO3}), hafnates, silicides expand beyond binaries.
\end{enumerate*}

Our predictions overlap prior ML/DFT efforts\cite{I18,kanai} on oxides and carbides but uniquely identify layered chalcogenides and complex oxides, with approximately 40\% of the candidate space (\ce{PbWO4}, \ce{HfS2}, \ce{CdWO4}, \ce{HfSnS3}, among others) not appearing in any prior screening study.  This complementarity positions the framework as an efficient composition-level pre-filter for targeted \textit{ab initio} and experimental validation.

\begin{table*}[htpb]
	\resizebox{.78\linewidth}{!}{
		\begin{tabular}{llccc|llccccl}\toprule
			Material ID & Formula        & $E_g$ (eV) & $\epsilon$ & Symmetry     & Material ID & Formula & $E_g$ (eV) & $\epsilon$ &
			Symmetry                                                                                                                  \\\midrule
			mp-1077316  & \ce{CO2}       & 7.655      & --         & $I\bar42d$   &
			mp-994911   & \ce{S}         & 3.327      & --         & $P4mm$                                                           \\
			mp-684668   & \ce{CaC2}      & 2.864      & --         & $Pnnm$       &
			mp-1000     & \ce{BaTe}      & 1.593      & 13.189     & $Fm\bar3m$                                                       \\
			mp-1018721  & \ce{HfO2}      & 4.668      & --         & $P4_2/nmc$   &
			mp-556068   & \ce{SiO2}      & 6.310      & --         & $Im\bar3m$                                                       \\
			mp-1204627  & \ce{Si}        & 1.522      & --         & $Cmcm$       &
			mp-755769   & \ce{ZrO2}      & 4.195      & 16.051     & $C2/m$                                                           \\
			mp-1101018  & \ce{TiZn2O4}   & 2.311      & --         & $P4_322$     &
			mp-7140     & \ce{SiC}       & 2.303      & 10.556     & $P6_3mc$                                                         \\
			mp-1245015  & \ce{ZnO}       & 1.276      & --         & $P1$         &
			mp-550714   & \ce{PbO}       & 2.230      & 10.674     & $Pca2_1$                                                         \\
			mp-13032    & \ce{MgS}       & 3.369      & 6.685      & $F\bar43m$   &
			mp-557719   & \ce{PbS}       & 2.096      & 125.581    & $Cmce$                                                           \\
			mp-4387     & \ce{SrZrO3}    & 3.612      & 42.799     & $Pnma$       &
			mp-1019544  & \ce{BaZrO3}    & 3.116      & --         & $I4/mcm$                                                         \\
			mp-1018722  & \ce{HgSe}      & 1.043      & --         & $P3_221$     &
			mp-13641    & \ce{TiCdO3}    & 2.906      & 23.728     & $R\bar3$                                                         \\
			mp-7826     & \ce{HgO}       & 1.310      & 10.245     & $P3_121$     &
			mp-1095294  & \ce{SiS2}      & 3.392      & --         & $P2_1/c$                                                         \\
			mp-669414   & \ce{HfPbO3}    & 2.779      & --         & $Pba2$       &
			mp-978844   & \ce{SrCa3O4}   & 3.372      & --         & $Pm\bar3m$                                                       \\
			mp-634      & \ce{HgS}       & 1.708      & 11.963     & $P3_121$     &
			mp-545622   & \ce{ZnWO4}     & 3.525      & --         & $C12/c1$                                                         \\
			mp-8725     & \ce{HfSnS3}    & 1.198      & 21.720     & $Pnma$       &
			mp-7487     & \ce{BaO}       & 2.768      & --         & $P4/nmm$                                                         \\
			mp-554134   & \ce{SnS}       & 1.668      & --         & $Aem2$       &
			mp-1039     & \ce{MgTe}      & 2.361      & 8.001      & $P6_3mc$                                                         \\
			mp-20244    & \ce{ZrPbS3}    & 1.247      & --         & $Pnma$       &
			mp-17275    & \ce{SrY2O4}    & 3.704      & --         & $Pnma$                                                           \\
			mp-553432   & \ce{TiO2}      & 3.421      & 8.704      & $P3_121$     &
			mp-1672     & \ce{CaS}       & 2.382      & 12.092     & $Fm\bar3m$                                                       \\
			mp-647557   & \ce{ZrPbO3}    & 3.209      & --         & $Pba2$       &
			mp-1265     & \ce{MgO}       & 4.429      & 10.772     & $Fm\bar3m$                                                       \\
			mp-1181264  & \ce{GeTe}      & 2.177      & --         & $Cm$         &
			mp-22009    & \ce{PbSe}      & 1.299      & --         & $Fmm2$                                                           \\
			mp-8484     & \ce{ZnO2}      & 2.158      & 10.796     & $Pa\bar3$    &
			mp-2605     & \ce{CaO}       & 3.630      & 16.471     & $Fm\bar3m$                                                       \\
			mp-569517   & \ce{C}         & 4.569      & --         & $R\bar3m$    &
			mp-2472     & \ce{SrO}       & 3.275      & 18.046     & $Fm\bar3m$                                                       \\
			mp-10497    & \ce{SrC2}      & 2.292      & --         & $C12/c1$     &
			mp-7812     & \ce{GeO2}      & 3.278      & 5.705      & $P4_12_12$                                                       \\
			mp-635713   & \ce{TiZnO3}    & 3.123      & --         & $P\bar1$     &
			mp-1253     & \ce{BaSe}      & 1.949      & 14.129     & $Fm\bar3m$                                                       \\
			mp-689140   & \ce{PbWO4}     & 3.622      & 27.323     & $P1$         &
			mp-8377     & \ce{TeO2}      & 3.060      & 16.163     & $P2_12_12_1$                                                     \\
			mp-569170   & \ce{CaTe}      & 1.632      & --         & $P6_3/mmc$   &
			mp-3378     & \ce{SrHfO3}    & 4.170      & 32.665     & $Pnma$                                                           \\
			mp-2758     & \ce{SrSe}      & 2.231      & 11.887     & $Fm\bar3m$   &
			mp-556576   & \ce{ZnS}       & 2.089      & --         & $R3m$                                                            \\
			mp-22147    & \ce{HfPbS3}    & 1.452      & --         & $Pnma$       &
			mp-19132    & \ce{HgWO4}     & 2.288      & --         & $C12/c1$                                                         \\
			mp-2526683  & \ce{WO3}       & 2.660      & --         & $Pm$         &
			mp-3771     & \ce{MgTiO3}    & 3.529      & 19.871     & $R\bar3$                                                         \\
			mp-8335     & \ce{Ba2ZrO4}   & 2.979      & 43.576     & $I4/mmm$     &
			mp-1190098  & \ce{CaHfS3}    & 1.524      & --         & $Pnma$                                                           \\
			mp-4571     & \ce{CaZrO3}    & 3.826      & 48.725     & $Pnma$       &
			mp-19039    & \ce{CdMoO4}    & 2.468      & 16.812     & $I4_1/a$                                                         \\
			mp-1079940  & \ce{Sr3CaO4}   & 3.294      & --         & $Pm\bar3m$   &
			mp-1500     & \ce{BaS}       & 2.149      & 14.953     & $Fm\bar3m$                                                       \\
			mp-1079882  & \ce{Ba3SrO4}   & 2.174      & --         & $Pm\bar3m$   &
			mp-2242     & \ce{GeS}       & 1.238      & 34.268     & $Pnma$                                                           \\
			mp-1958     & \ce{SrTe}      & 1.762      & 12.399     & $Fm\bar3m$   &
			mp-1192077  & \ce{BaC2}      & 2.915      & --         & $Pnma$                                                           \\
			mp-1192803  & \ce{Lu2PbS4}   & 2.302      & --         & $Pnma$       &
			mp-7607     & \ce{TiHgO3}    & 1.094      & --         & $R\bar3c$                                                        \\
			mp-559545   & \ce{SeO2}      & 3.442      & --         & $Pmc2_1$     &
			mp-569679   & \ce{ZnSe}      & 1.967      & --         & $P4/nmm$                                                         \\
			mp-1208643  & \ce{SrHfS3}    & 1.496      & 45.049     & $Pnma$       &
			mp-984729   & \ce{BaSr3O4}   & 2.520      & --         & $Pm\bar3m$                                                       \\
			mp-1078644  & \ce{SnO}       & 1.643      & --         & $Cmc2_1$     &
			mp-2604     & \ce{MgTe2}     & 1.116      & 14.773     & $Pa\bar3$                                                        \\
			mp-985829   & \ce{HfS2}      & 1.224      & 26.671     & $P\bar3m1$   &
			mp-1008223  & \ce{CaSe}      & 3.166      & 6.428      & $F\bar43m$                                                       \\
			mp-3915     & \ce{BaHgO2}    & 2.273      & 8.382      & $R\bar3m$    &
			mp-1041984  & \ce{SnO2}      & 2.089      & 7.386      & $P6_3/mmc$                                                       \\
			mp-13066    & \ce{Er2O3}     & 4.197      & --         & $P\bar3m1$   &
			mp-1383750  & \ce{MgTi2O5}   & 3.078      & --         & $Pmmn$                                                           \\
			mp-1087     & \ce{SrS}       & 2.497      & 11.869     & $Fm\bar3m$   &
			mp-19363    & \ce{HgMoO4}    & 2.256      & 12.777     & $C12/c1$                                                         \\
			mp-3952     & \ce{BaY2O4}    & 3.252      & --         & $Pnma$       &
			mp-1186     & \ce{ZrS2}      & 1.042      & 35.900     & $P\bar3m1$                                                       \\
			mp-1021511  & \ce{CdS}       & 1.519      & 5.5980     & $P3m1$       &
			mp-20078    & \ce{Pb2O3}     & 1.075      & 27.991     & $P2_1/c$                                                         \\
			mp-27221    & \ce{Ca2Hf7O16} & 4.305      & --         & $R\bar3$     &
			mp-22681    & \ce{Pb2WO5}    & 3.089      & --         & $C2/m$                                                           \\
			mp-5548     & \ce{Ta2Cd2O7}  & 2.027      & 176.467    & $Fd-3m$      &
			mp-1105549  & \ce{BaHfS3}    & 1.252      & 58.187     & $Pnma$                                                           \\
			mp-21452    & \ce{Pb3O4}     & 1.160      & --         & $Pbam$       &
			mp-1106215  & \ce{TiPbO3}    & 2.316      & --         & $I4/m$                                                           \\
			\bottomrule
		\end{tabular}
	}
	\caption{Entries from the Materials Project database predicted by EC at $>0.95$ confidence using the constrained voting
		strategy, filtered to retain only compositions with PBE bandgap $E_g \geq$  1.0 eV. Of the 122 candidates identified by the
		ensemble, 32 with $E_g$ < 1.0 eV are omitted since wide-bandgap phases are more likely to host optically addressable spin states
		in the defects. The ML confidence score measures compositional compatibility with known hosts; a secondary electronic-structure
		filter remains a necessary downstream step. All relevant properties are retrieved from the Materials Project
		database using the MPRester API\cite{ong2015materials}. In case of polymorphism, the phase with the widest bandgap is
		chosen provided that the material is thermodynamically stable with energy above the convex hull $E_{\text{Hull}} <
			0.2$\,eV/atom.}
	\label{tab:predictions}
\end{table*}

\subsection{First-Principles Validation of Dielectric Screening and Defect Physics}
\label{sec:validation}

\begin{figure*}[htpb]
	\centering
	\begin{minipage}[b]{.55\textwidth}
		\centering
		\begin{tabularx}{.65\textwidth}{lllHll}
			\toprule
			Material   & $\varepsilon_{{elec}}$ & $\varepsilon_{{ionic}}$ & $\varepsilon_{{total}}$ & $ E_g$(eV) & $T_2$(ms)              \\
			\midrule
			WS$_2$     & 4.22$^{*}$             & 0.01$^{*}$              & 4.23$^{*}$              & 1.81       & 11\cite{I10}           \\
			SiC        & 7.20                   & 3.43                    & 10.63                   & 2.30       & 1.1\cite{I15SiCQubits} \\
			CaS        & 5.01                   & 7.25                    & 12.26                   & 2.38       & 23\cite{kanai}         \\
			CaO        & 3.78                   & 12.65                   & 16.43                   & 3.63       & 34\cite{C2}            \\
			MgO        & 3.18                   & 6.37                    & 9.55                    & 4.43       & 0.6\cite{C2}           \\
			C          & 5.83                   & 0.00                    & 5.83                    & 4.57       & 0.89\cite{I15NVCenter} \\
			CaCO$_3$   & 2.72                   & 5.59                    & 8.31                    & 5.00       & 11\cite{kanai}         \\
			\midrule
			\ce{HfS2}  & 8.49                   & 24.80                   & 33.29                   & 1.22       & $\dagger$              \\
			\ce{CaTe}  & 6.76                   & 6.83                    & 13.59                   & 1.63       & 17.95                  \\
			\ce{BaO}   & 4.58                   & 11.34                   & 15.92                   & 2.77       & 28.51                  \\
			\ce{TiO2}  & 7.20                   & 48.12                   & 55.32                   & 3.42       & $\dagger$              \\
			\ce{PbWO4} & 4.79$^{*}$             & 23.80$^{*}$             & 28.59$^{*}$             & 3.62       & $\dagger$              \\
			\bottomrule
			\multicolumn{6}{c}{\textit{\small $\ast$ Averaged over anisotropic tensor components}}                                        \\
		\end{tabularx}

		\captionof{table}{Static dielectric properties and bandgaps calculated via DFT (DFPT) for 12 validation materials.
			Top section: Known quantum hosts with $T_2$ values probed from literature, compared to $\varepsilon$ from
			the present work. Bottom section: Novel predictions with $T_2$ estimated from empirical scaling $T_2 =
				4\varepsilon^{3.269} \times 10^{-3}$ ms (see \figref{fig:dielectric_correlation}).
			For \ce{CaTe} and \ce{BaO}, which fall within the fitting range ($\varepsilon_{\text{total}} \approx 4\text{--}17$), predicted
			$T_2$ values are order-of-magnitude estimates and require experimental validation. $\dagger$\ce{TiO2}, $\dagger$\ce{HfS2} and
			$\dagger$\ce{PbWO4} lie outside the fitting range; no scaled $T_2$ is reported as the extrapolation is unreliable. The
			fitted exponent (3.269) is an empirical parameter without independent theoretical derivation.
			Further, dominant ionic contribution in rutile arises partly from soft phonon modes, which can also act
			as decoherence channels and are not captured by the dielectric screening proxy. The promise of \ce{TiO2} as a
			quantum host rests instead on its deep mid-gap defect states, near-complete nuclear-spin-free lattice, and structural analogy to \ce{MgO} and \ce{CaO} (see text).}
		\label{tab:dft}
	\end{minipage}\hfill
	\begin{minipage}[b]{.42\textwidth}
		\centering
		\includegraphics[width=\textwidth]{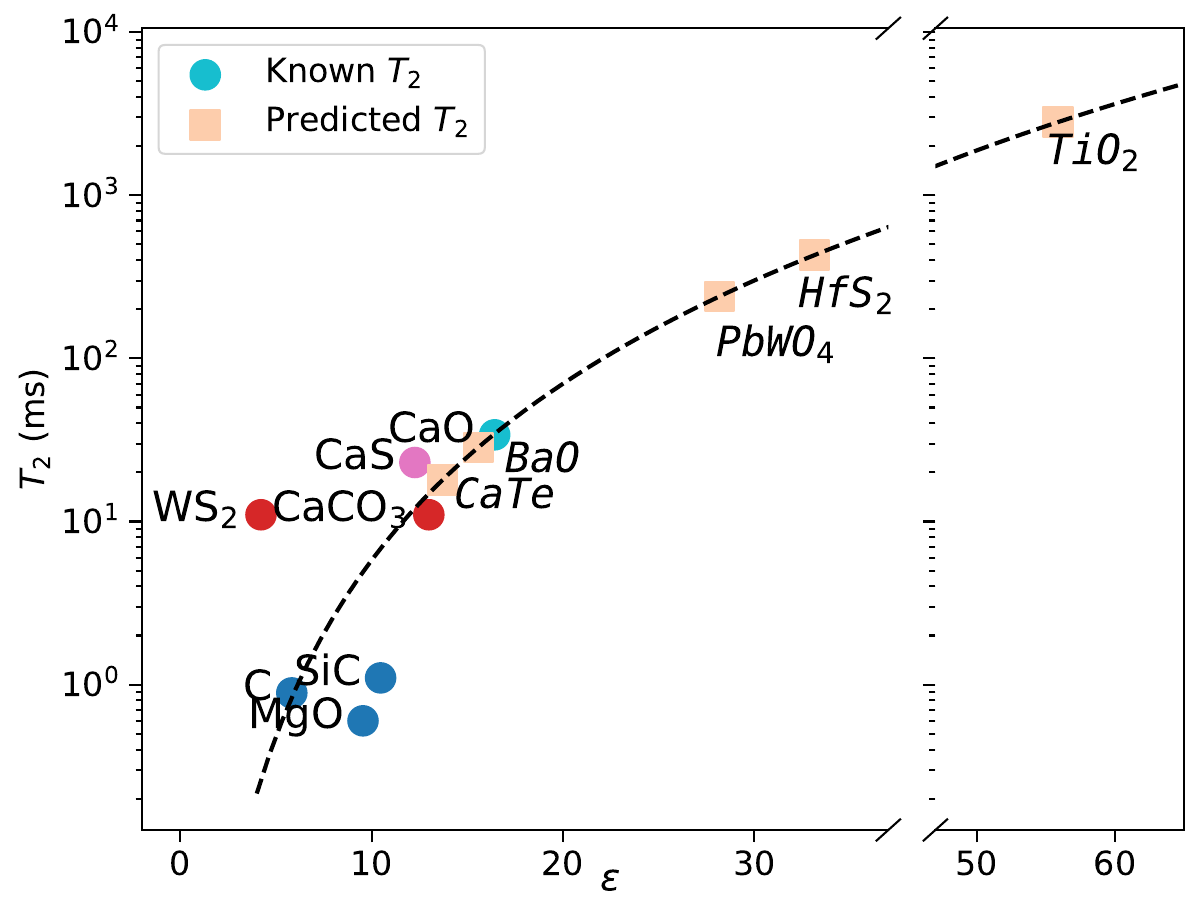}
		\captionof{figure}{Correlation between DFT-calculated static dielectric constants ($\varepsilon$, this work) and spin
			coherence times ($T_2$). Circles: experimentally measured $T_2$ from literature. Squares: $T_2$ estimated from
			empirical scaling $T_2 = 4\varepsilon^{3.269} \times 10^{-3}$\,ms, derived from and applied within the fitting range
			of known hosts ($\varepsilon_{\text{total}} \approx 4$--17). The strong correlation ($R^2 = 0.89$) validates dielectric
			screening as a robust proxy for coherence within this chemical family. \ce{HfS2} and \ce{PbWO4} lie near the upper
			edge of the fitting range and their estimated $T_2$ values are shown as indicative. \ce{TiO2}
			($\varepsilon_{\text{total}} = 55.32$) falls well outside the fitting range and no scaled $T_2$ is reported; its
			candidacy is evaluated on independent physical grounds (see text and \tabref{tab:dft}).\vspace{1cm}}
		\label{fig:dielectric_correlation}
	\end{minipage}
\end{figure*}

To assess the physical viability of the ML-predicted candidates and to ground the screening in first-principles physics, we performed density functional perturbation theory (DFPT) calculations of the static dielectric tensor ($\varepsilon_{\text{total}}$) for 12 representative materials spanning five distinct chemical families (\tabref{tab:dft}).  The validation set consists of (i) seven experimentally established quantum hosts (diamond, SiC, \ce{WS2}, CaS, CaO, MgO, \ce{CaCO3}), which serve as internal benchmarks for the DFT methodology, and (ii) five novel predictions selected to maximize chemical diversity: simple binary oxides (\ce{TiO2}, BaO), a layered chalcogenide (\ce{HfS2}), an alkaline earth telluride (CaTe), and a complex oxide (\ce{PbWO4}).

The dielectric constant plays a central role in suppressing spin decoherence by screening fluctuating electric
fields\cite{M1,nagura2026understanding,I17}.  As shown in \figref{fig:dielectric_correlation}, our calculated
$\varepsilon_{\text{total}}$ values for the seven known hosts correlate strongly with known $T_2$ coherence times ($R^2 = 0.89$).
The resulting empirical scaling $T_2 \propto \varepsilon^{3.27}$ provides a quantitative estimate of $T_2$ in the limiting case
of a nuclear-spin free "silent lattice".
We note that the scaling provides upper-bound estimates of coherence times, as additional decoherence channels---nuclear-spin
noise, defect-symmetry-dependent spin--orbit coupling, and phonon-mediated relaxation---are not captured by the dielectric proxy
alone.

\subsubsection{Benchmarking Against Known Quantum Hosts}

The DFPT-calculated dielectric tensors for the seven established hosts reproduce the expected diversity of screening mechanisms across bonding classes.  Diamond exhibits a purely electronic response ($\varepsilon_{\text{total}} = 5.83$, $\varepsilon_{\text{ionic}} = 0$), consistent with its rigid covalent lattice and the absence of infrared-active phonon modes.  In contrast, the rocksalt oxides CaO and MgO display dominant ionic contributions, with $\varepsilon_{\text{ionic}}/\varepsilon_{\text{total}}$ ratios of 0.77 and 0.67, respectively, reflecting the strong Born effective charges of the alkaline-earth--oxygen bond.  The layered TMDC \ce{WS2} shows a comparatively small ionic component ($\varepsilon_{\text{total}} = 4.23$, $\varepsilon_{\text{ionic}} \approx 0.01$), consistent with weak interlayer bonding and predominantly covalent in-plane character.  The intermediate cases---SiC (mixed covalent--ionic), CaS (softer anion), and \ce{CaCO3} (molecular anion)---fall between these extremes, and our calculated values agree with the literature to within 5--10\%, validating the DFPT methodology and providing confidence in extending it to previously uncharacterized compounds.

\subsubsection{Novel Predictions Across Chemical Families}

The five novel candidates illustrate three physically distinct pathways to achieving high dielectric screening:

\begin{enumerate}[label=(\roman*)]
	\item \textbf{Simple binary oxides:} \ce{TiO2} exhibits the highest dielectric constant among all studied materials
	      ($\varepsilon_{\text{total}} = 55.32$), with a dominant ionic contribution ($\varepsilon_{\text{ionic}} = 48.12$) arising from strong lattice polarizability associated with the Ti--O stretching modes.
	      BaO shows $\varepsilon_{\text{total}} = 15.92$ ($\varepsilon_{\text{ionic}} = 11.34$), placing it in the same screening regime as the known oxide hosts CaO and MgO.

	\item \textbf{Layered chalcogenides:} \ce{HfS2} demonstrates remarkably strong screening ($\varepsilon_{\text{total}} =
		      33.29$), nearly an order of magnitude larger than the isostructural \ce{WS2} ($\varepsilon_{\text{total}} = 4.23$). This enhancement arises primarily from the greater ionic polarizability of the Hf--S bond relative to the more covalent W--S bond, as reflected in the ionic ratio $\varepsilon_{\text{ionic}}/\varepsilon_{\text{total}} = 0.74$ for \ce{HfS2} versus $< 0.01$ for \ce{WS2}.  CaTe ($\varepsilon_{\text{total}} = 13.59$) confirms that substantial screening extends across the heavier chalcogenides.

	\item \textbf{Complex oxides:} \ce{PbWO4} exhibits a balanced electronic--ionic response ($\varepsilon_{\text{elec}} = 4.79$, $\varepsilon_{\text{ionic}} = 23.80$), yielding $\varepsilon_{\text{total}} = 28.59$.  The scheelite structure accommodates both the stereochemically active Pb 6$s^2$ lone pair and the rigid WO$_4^{2-}$ tetrahedra, producing a screening profile not present in simpler binary hosts and suggesting that complex oxide families may offer an underexplored design space for quantum host engineering.
\end{enumerate}

Applying the empirical $T_2(\varepsilon)$ scaling to candidates within the fitting range
($\varepsilon_{\text{total}} \approx 4$--17) yields estimated coherence times of approximately 8 ms for \ce{CaTe} and 29 ms for
\ce{BaO} both substantially exceeding diamond's $T_2 \approx 0.89$~ms\cite{I15NVCenter}.  These values provide a
relative ranking of candidates rather than quantitative predictions: the scaling is derived from only seven data points spanning
a limited chemical range, and the fitted exponent (3.269) is an empirical parameter without an independently derived theoretical
basis. \ce{HfS2} ($\varepsilon$ = 33.29) and \ce{PbWO4} ($\varepsilon$ = 28.59) lie well beyond the upper edge of the fitting
range ($\varepsilon \approx 17$ for CaO); we do not report extrapolated $T_2$ values for these materials, as the reliability of
the scaling cannot be assessed at these dielectric constants. Their candidacy rests on the magnitude of their dielectric
screening rather than on numerically estimated coherence times.

\ce{TiO2}, with $\varepsilon_{\text{total}} = 55.32$ lies more than three times beyond
the upper edge of the fitting range, making any extrapolated $T_2$ physically meaningless.  More fundamentally, the anomalously
large ionic dielectric response of rutile \ce{TiO2} is driven in part by soft phonon modes near the zone boundary---the same
lattice dynamics that can mediate electric-field-induced decoherence\cite{M1, I17}.  A proxy that treats high
$\varepsilon_{\text{ionic}}$ as uniformly beneficial cannot account for this dual role.  The case for \ce{TiO2} as a quantum host
accordingly rests on independent physical grounds, which we develop in the following subsection.

\subsubsection{Defect Energetics and Electronic Structure}

\begin{figure}[htpb]
	\centering
	\begin{subfigure}[b]{.45\textwidth}
		\begin{center}
			\includegraphics[width=.95\textwidth]{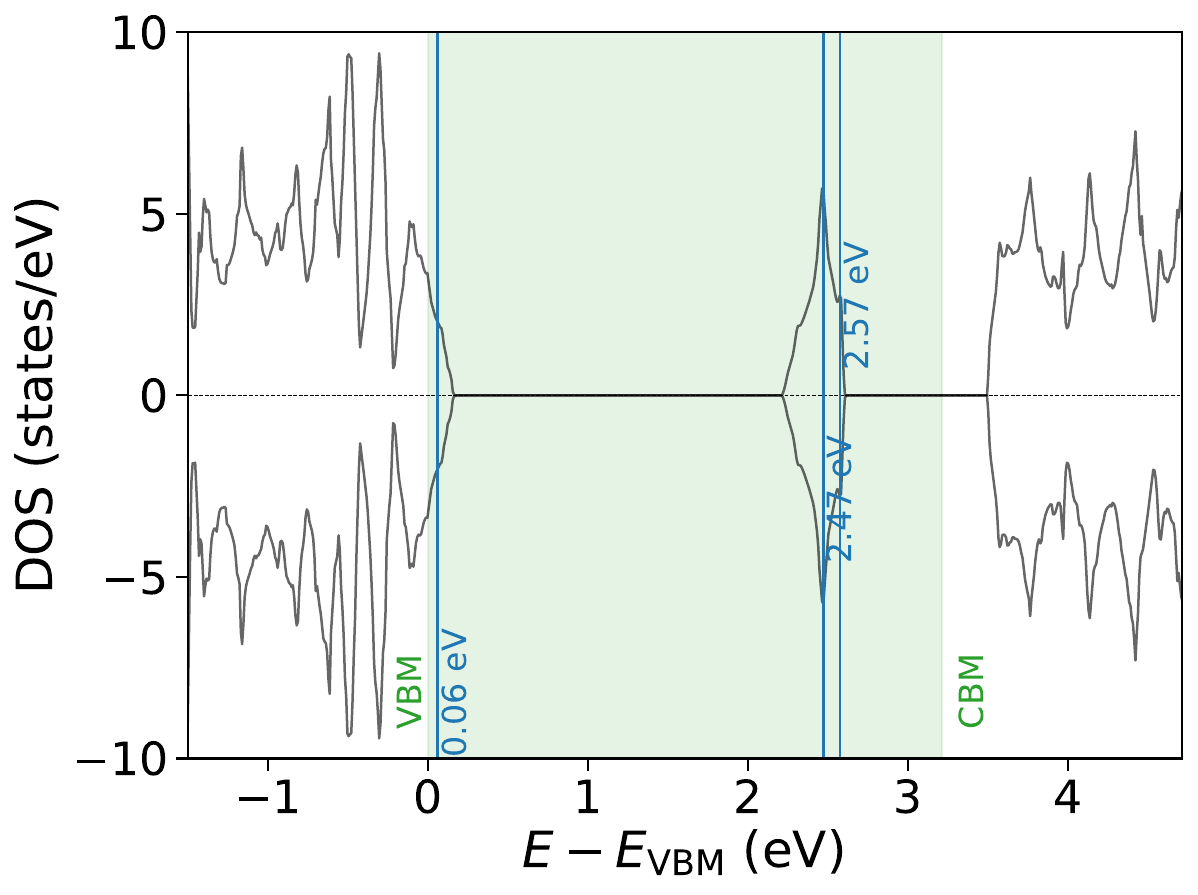}
		\end{center}
		\caption{}
		\label{fig:defect_tio2}
	\end{subfigure}
	\begin{subfigure}[b]{.45\textwidth}
		\begin{center}
			\includegraphics[width=.95\textwidth]{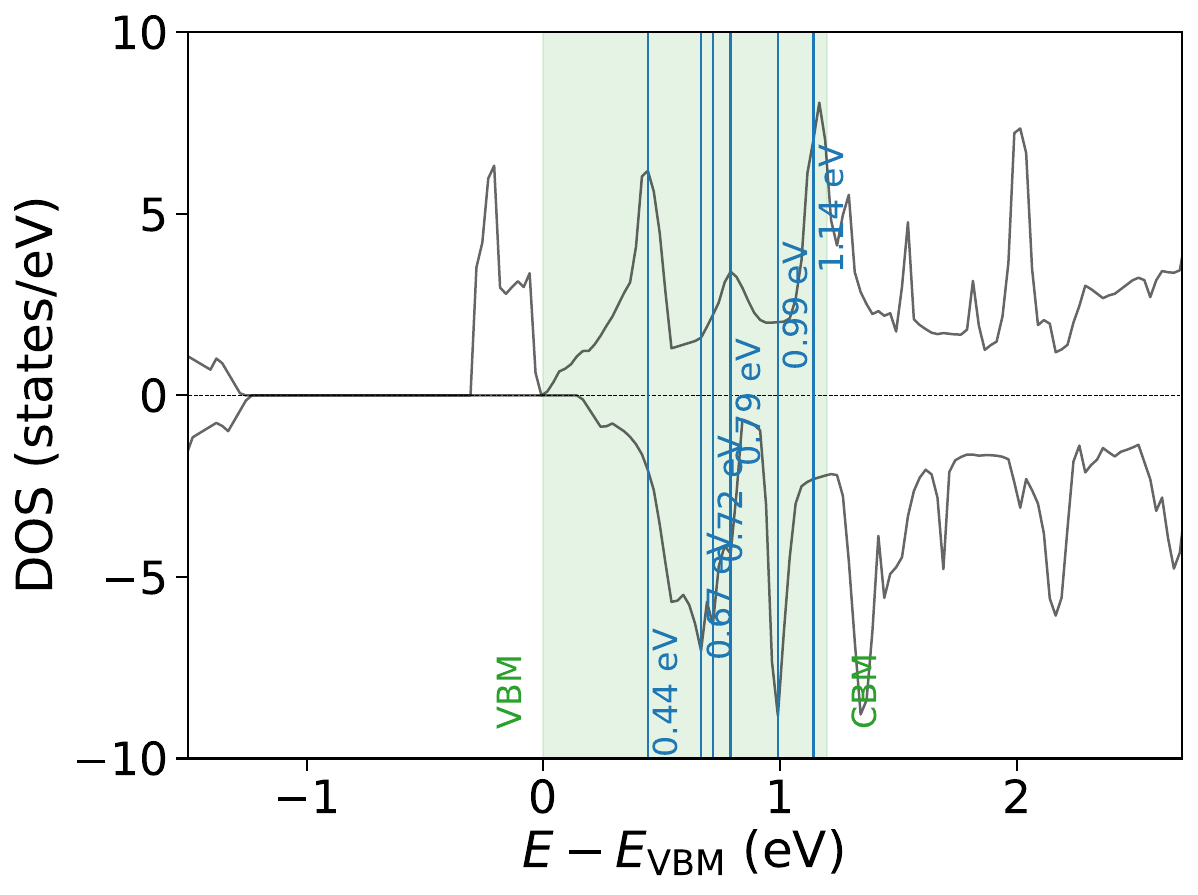}
		\end{center}
		\caption{}
		\label{fig:defect_hfs2}
	\end{subfigure}

	\caption{Defect density of states for (\subref{fig:defect_tio2}) oxygen-deficient \ce{TiO2} and (\subref{fig:defect_hfs2}) sulfur-deficient \ce{HfS2}, calculated at the PBE level using $2\times 2\times 2$ supercells with FNV charge corrections.  In \ce{TiO2}, the oxygen vacancy introduces deep, well-isolated mid-gap states centered between 2.4 and 2.65~eV above the VBM, separated from both band edges by $>0.7$~eV---a configuration favorable for optical addressability and spin-state isolation.  In \ce{HfS2}, the sulfur vacancy produces several states distributed more broadly within the narrower 1.2~eV gap, with the dominant peak at $\sim$0.6~eV above the VBM.  Bandgap underestimation inherent to PBE may affect the absolute positioning of these levels; hybrid-functional calculations are needed to refine the defect-level placement in both materials.}
	\label{fig:dielectric_correlation_1}
\end{figure}

To complement the dielectric analysis, we examined vacancy-induced defect states in two representative materials: \ce{TiO2} (the highest-$\varepsilon$ oxide) and \ce{HfS2} (a high-screening layered chalcogenide).  The defect densities of states are shown in \figref{fig:defect_tio2} and \figref{fig:defect_hfs2}.

In \ce{TiO2}, the neutral oxygen vacancy produces several deep, well-localized in-gap states.  The dominant peak
lies 2.48~eV above the valence band maximum (VBM) and 0.73~eV below the conduction band minimum (CBM), with a high local
density of states (11.57~states/eV).  The isolated, mid-gap character of these states is structurally favorable for charge
localization and optical addressability---prerequisites for spin-defect-based quantum
applications\cite{awschalom_quantum_2018}.  When combined with its host-lattice attributes---a near-complete nuclear-spin-free
Ti sublattice ($\approx 90\%$ spin-zero isotopes: $^{46}$Ti, $^{48}$Ti, $^{50}$Ti), well-established protocols for isotopic
purification of oxygen in thin films\cite{jeong2022strong}, CMOS-compatible synthesis, and decades of native defect engineering
in rutile and anatase\cite{gilliard2017manipulation}---\ce{TiO2} emerges as a compelling candidate on a trajectory analogous to
that of \ce{MgO} and \ce{CaO} at earlier stages of their development as qubit hosts\cite{C2}.  The soft-phonon-driven ionic
polarizability that renders the dielectric scaling inapplicable does not preclude long coherence; rather, it identifies phonon
engineering---through strain, doping, or choice of polymorph---as the key experimental lever for realizing the material's
potential.

In \ce{HfS2}, the sulfur vacancy gives rise to multiple states distributed across the narrower 1.2~eV bandgap.  The most
prominent peak appears 0.58~eV above the VBM (0.62~eV below the CBM) with a lower density ($\sim$6~states/eV) than the \ce{TiO2}
oxygen vacancy. While the proximity of these states to the band edges at the PBE level of theory suggests a
greater susceptibility to thermal ionization and phonon-assisted transitions, this comparison should be interpreted with
caution: the PBE functional systematically underestimates bandgaps, and hybrid-functional or GW corrections would likely shift
the defect levels deeper into the gap.  What can be concluded from the present calculations is that (i) the sulfur vacancy does
introduce in-gap states, confirming that \ce{HfS2} can host point defects with electronic signatures, and (ii) the defect
landscape is qualitatively different from that of \ce{TiO2}, with more dispersed states that warrant higher-level electronic
structure characterization before coherence properties can be assessed.  The exceptionally high dielectric screening of
\ce{HfS2} ($\varepsilon_{\text{total}} = 33.29$) may partially compensate for the less favorable defect positioning by
suppressing charge-noise-induced decoherence, but this tradeoff requires explicit spin-dynamics modeling to quantify.

\subsubsection{Validation Summary}

The 12-material first-principles analysis demonstrates:
\begin{enumerate*}[label=(\roman*)]
	\item quantitative agreement between DFPT-calculated dielectric tensors and established screening behavior across four distinct bonding classes (covalent, ionic, mixed, and van der Waals),
	\item a strong correlation ($R^2 = 0.89$) between $\varepsilon_{\text{total}}$ and experimental $T_2$ that validates dielectric screening as a coherence proxy within the studied chemical range,
	\item physically distinct pathways to high screening in novel candidates---lattice polarizability (\ce{TiO2}), heavy-cation ionic response (\ce{HfS2}), and balanced electronic--ionic contributions (\ce{PbWO4})---none of which were represented among prior experimentally explored hosts, and
	\item defect electronic structures consistent with spin-defect hosting in both \ce{TiO2} (deep, isolated mid-gap states) and \ce{HfS2} (in-gap states requiring further refinement at higher levels of theory).
\end{enumerate*}

This validation encompasses $\sim$10\% of the 122 high-confidence predictions (\tabref{tab:predictions}) and spans the major
chemical families represented therein.  The agreement between ML-predicted compatibility and DFT-calculated physical properties
confirms that the composition-only ensemble captures transferable screening physics rather than dataset-specific correlations.
The remaining candidates constitute a prioritized target list for experimental synthesis and defect engineering, with layered
chalcogenides (\ce{HfS2}, \ce{SnS2}, \ce{TiS2}, \ce{ZrS2}) especially promising for nanoscale quantum device integration via
established exfoliation methods, and high-$\varepsilon$ oxides (\ce{TiO2}, \ce{BaO}, tungstates) as bulk quantum memory
platforms.

\section{Conclusion}

This work demonstrates that composition-only machine learning, when built on Rashomon set ensembles and interrogated with
model-agnostic explainability tools, can simultaneously screen the quantum defect host landscape at scale and extract the
physical principles governing that landscape.  Screening $\sim$45,000 thermodynamically stable compounds from the Materials
Project in minutes, the framework identifies 122 high-confidence candidates (confidence $>0.95$) that recover most experimentally
verified quantum host---diamond, \ce{SiC}, \ce{ZnO}, \ce{ZnS}, and, at reduced confidence, the layered TMDCs \ce{WS2} and
\ce{MoS2}---while predicting previously unexplored materials across oxides (\ce{TiO2}, \ce{BaO}, tungstates), layered
chalcogenides (\ce{HfS2}, \ce{ZrS2}, \ce{SnS2}), alkaline earth tellurides, and complex ternaries.  DFPT validation of 12
representative materials confirms that the ML-predicted compatibility is grounded in real screening physics ($R^2 = 0.89$ between
$\varepsilon_{\text{total}}$ and experimental $T_2$), and defect-level calculations identify deep mid-gap vacancy states in
\ce{TiO2} that, combined with its near-complete nuclear-spin-free lattice and CMOS-compatible synthesis, position it as the most
compelling novel candidate to emerge from this study.

The principal intellectual contribution, however, is not the candidate list itself but the method and the design rules distilled from it.  By
contrasting the feature attributions of seven diverse classifiers aggregated into a heterogeneous ensemble, the Rashomon+ALE
analysis reveals consensus physical criteria that no single model identifies alone: compositions with low chemical heterogeneity
and filled $s$-, $d$-, and $f$-shells are strongly favored; enrichment in C, S, Si, and O increases predicted compatibility while
Co, V, P, Al, and Mn suppress it; and wider atomic-orbital-derived bandgaps correlate positively with quantum host suitability.
These rules---interpretable, transferable, and consistent with independent \textit{ab initio} screening\cite{kanai}---move beyond a prioritized to-do list toward a physically grounded grammar for rational defect host engineering that can guide materials design in chemical spaces not represented in the training data\cite{MFE5}.

The framework has a clear and deliberate limitation.  Because the feature space encodes only composition and the
positive-class filter requires stable spin-zero isotopes, the model systematically excludes hosts in which coherence is an
emergent property of structural engineering---most notably \ce{hBN} and \ce{GaN}, where long $T_2$ times are achieved through
dimensionality reduction and isotopic purification of intrinsically spinful lattices.  This is not a failure of the model but a
boundary condition of the descriptor choice: the composition-only approach trades coverage of structure-dependent coherence
mechanisms for the ability to screen the entire chemical space without the computational burden of structural relaxation.
Extending the framework to capture such mechanisms would require incorporating structural descriptors (e.g., layer thickness,
interlayer coupling, phonon spectra) at the cost of the current screening speed, a tradeoff that merits exploration in future
work.

Looking ahead, the 110 unvalidated predictions define a concrete experimental and computational roadmap.  The immediate priority is automated DFT workflows---using hybrid functionals (HSE06) and spin--orbit coupling---to map defect formation energies, charge transition levels, and hyperfine parameters across the layered chalcogenide family (\ce{HfS2}, \ce{SnS2}, \ce{TiS2}, \ce{ZrS2}), which are most accessible to near-term experimental validation via established exfoliation methods.  In parallel, thin-film synthesis, ion implantation, and spin resonance measurements on \ce{TiO2} and the high-$\varepsilon$ oxides (\ce{BaO}, \ce{SrO}, tungstates, molybdates) would test the dielectric screening proxy in bulk quantum memory architectures.  As experimental feedback accumulates, iterative retraining will sharpen the ensemble's decision boundary and may reveal additional design rules not yet resolvable from the current training set.  More broadly, the Rashomon+XAI methodology demonstrated here is not specific to quantum defect hosts: any materials discovery problem in which multiple competing physical criteria must be balanced---single-photon emitters, radiation-tolerant materials, solid-state battery electrolytes---could benefit from the same approach of extracting consensus design rules from ensembles of disagreeing models.

\section*{Data Availability}
\label{sec:data_availability}
The data used in this study is available at this GitHub repository\cite{mahshook2025}.
\section*{Author contributions}
\label{sec:author_contributions}
Mohammed Mahshook: Conceptualization, Methodology, Data Curation, Formal Analysis, Software, Visualization, Writing – Original Draft Preparation.
Rudra Banerjee: Conceptualization, Methodology, Formal Analysis, Resources, Supervision, Validation, Writing – Review \& Editing

\section*{Funding}
Not applicable

\section*{Competing interests}
The authors declare no competing interests.


\begin{thebibliography}{10}

	\bibitem{I21QuantumReadiness}
	Abhishek Purohit, Maninder Kaur, Zeki~Can Seskir, Matthew~T Posner, and Araceli Venegas-Gomez.
	\newblock Building a quantum-ready ecosystem.
	\newblock {\em IET Quantum Communication}, 5(1):1--18, 2024.

	\bibitem{divincenzo2025thirty}
	David~P DiVincenzo.
	\newblock Thirty years of quantum computing.
	\newblock {\em Quantum Science and Technology}, 10(3):030501, 2025.

	\bibitem{I2QCReview}
	Pharnam Bakhshinezhad, Mohammad Mehboudi, Carles Roch~I Carceller, and Armin Tavakoli.
	\newblock Scalable entanglement certification via quantum communication.
	\newblock {\em PRX Quantum}, 5(2):020319, 2024.

	\bibitem{I3QCReview}
	David DeMille, Nicholas~R Hutzler, Ana~Maria Rey, and Tanya Zelevinsky.
	\newblock Quantum sensing and metrology for fundamental physics with molecules.
	\newblock {\em Nature Physics}, 20(5):741--749, 2024.

	\bibitem{kaur2025neutral}
	Inderpreet Kaur, Neha Singh, Suranita Kanjilal, Bodhaditya Santra, et~al.
	\newblock Neutral atom quantum computers for the applications in condensed matter physics.
	\newblock {\em Journal of Physics: Condensed Matter}, 37(17):173001, 2025.

	\bibitem{I6SQubits}
	John Clarke and Frank~K Wilhelm.
	\newblock Superconducting quantum bits.
	\newblock {\em Nature}, 453(7198):1031--1042, 2008.

	\bibitem{I6PQubits}
	Srujan Meesala, David Lake, Steven Wood, Piero Chiappina, Changchun Zhong, Andrew~D Beyer, Matthew~D Shaw, Liang Jiang, and Oskar Painter.
	\newblock Quantum entanglement between optical and microwave photonic qubits.
	\newblock {\em Physical Review X}, 14(3):031055, 2024.

	\bibitem{IE5}
	John Preskill.
	\newblock Quantum computing in the nisq era and beyond.
	\newblock {\em Quantum}, 2:79, 2018.

	\bibitem{awschalom_quantum_2018}
	DD~Awschalom, R~Hanson, J~Wrachtrup, and BB~Zhou.
	\newblock Quantum technologies with optically interfaced solid-state spins.
	\newblock {\em Nature Photonics}, 12:516--527, 2018.

	\bibitem{seo2017designing}
	Hosung Seo, He~Ma, Marco Govoni, and Giulia Galli.
	\newblock Designing defect-based qubit candidates in wide-gap binary semiconductors for solid-state quantum technologies.
	\newblock {\em Physical Review Materials}, 1(7):075002, 2017.

	\bibitem{I14NVCenter}
	NB~Manson, JP~Harrison, and MJ~Sellars.
	\newblock Nitrogen-vacancy center in diamond: Model of the electronic structure and associated dynamics.
	\newblock {\em Physical Review B—Condensed Matter and Materials Physics}, 74(10):104303, 2006.

	\bibitem{I15NVCenter}
	Laura Orphal-Kobin, Cem~G{\"u}ney Torun, Julian~M Bopp, Gregor Pieplow, and Tim Schr{\"o}der.
	\newblock Coherent microwave, optical, and mechanical quantum control of spin qubits in diamond.
	\newblock {\em Advanced Quantum Technologies}, 8(2):2300432, 2025.

	\bibitem{Atature_2018}
	Mete Atatüre, Dirk Englund, Nick Vamivakas, Sang-Yun Lee, and Joerg Wrachtrup.
	\newblock Material platforms for spin-based photonic quantum technologies.
	\newblock {\em Nature Reviews Materials}, 3(5):38–51, apr 2018.

	\bibitem{wolfowicz_quantum_2021}
	G~Wolfowicz, FJ~Heremans, and CP~et~al. Anderson.
	\newblock Quantum guidelines for solid-state spin defects.
	\newblock {\em Nature Reviews Materials}, 6:906--925, 2021.

	\bibitem{I15SiCQubits}
	Stephanie Simmons.
	\newblock Scalable fault-tolerant quantum technologies with silicon color centers.
	\newblock {\em PRX quantum}, 5(1):010102, 2024.

	\bibitem{carbone2025quantifying}
	Amedeo Carbone, Ilia~D Breev, Johannes Figueiredo, Silvan Kretschmer, Leonard Geilen, Amine Ben~Mhenni, Johannes Arceri, Arkady~V Krasheninnikov, Martijn Wubs, Alexander~W Holleitner, et~al.
	\newblock Quantifying the creation of negatively charged boron vacancies in he-ion irradiated hexagonal boron nitride.
	\newblock {\em Physical Review Materials}, 9(5):056203, 2025.

	\bibitem{I10}
	Song Li, Gerg{\H{o}} Thiering, P{\'e}ter Udvarhelyi, Viktor Iv{\'a}dy, and Adam Gali.
	\newblock Carbon defect qubit in two-dimensional ws2.
	\newblock {\em Nature communications}, 13(1):1210, 2022.

	\bibitem{I16TMDQubits}
	Yeonghun Lee, Yaoqiao Hu, Xiuyao Lang, Dongwook Kim, Kejun Li, Yuan Ping, Kai-Mei~C Fu, and Kyeongjae Cho.
	\newblock Spin-defect qubits in two-dimensional transition metal dichalcogenides operating at telecom wavelengths.
	\newblock {\em Nature Communications}, 13(1):7501, 2022.

	\bibitem{M1}
	Gary Wolfowicz, F~Joseph Heremans, Christopher~P Anderson, Shun Kanai, Hosung Seo, Adam Gali, Giulia Galli, and David~D Awschalom.
	\newblock Quantum guidelines for solid-state spin defects.
	\newblock {\em Nature Reviews Materials}, 6(10):906--925, 2021.

	\bibitem{IE8}
	JR~Weber, WF~Koehl, JB~Varley, Anderson Janotti, BB~Buckley, CG~Van~de Walle, and David~D Awschalom.
	\newblock Quantum computing with defects.
	\newblock {\em Proceedings of the National Academy of Sciences}, 107(19):8513--8518, 2010.

	\bibitem{I17}
	Austin~M Ferrenti, Nathalie~P de~Leon, Jeff~D Thompson, and Robert~J Cava.
	\newblock Identifying candidate hosts for quantum defects via data mining.
	\newblock {\em npj Computational Materials}, 6(1):126, 2020.

	\bibitem{I12}
	DW~Davies, KT~Butler, AJ~Jackson, A~Morris, JM~Frost, JM~Skelton, and A~Walsh.
	\newblock Computational screening of all stoichiometric inorganic materials. chem 2016, 1 (4), 617--627.

	\bibitem{Seo_2016}
	Hosung Seo, Abram~L. Falk, Paul~V. Klimov, Kevin~C. Miao, Giulia Galli, and David~D. Awschalom.
	\newblock Quantum decoherence dynamics of divacancy spins in silicon carbide.
	\newblock {\em Nature Communications}, 7(1), sep 2016.

	\bibitem{ivady_ab_2019}
	V~Iv{\'a}dy, J~Davidsson, NTT Son, and et~al.
	\newblock Ab initio theory of the diamond nv center as a quantum sensor for strain and electric field.
	\newblock {\em npj Quantum Materials}, 4:1--8, 2019.

	\bibitem{IE9}
	Rodrick Kuate~Defo, Haimi Nguyen, Mark~JH Ku, and Trevor~David Rhone.
	\newblock Methods to accelerate high-throughput screening of atomic qubit candidates in van der waals materials.
	\newblock {\em Journal of Applied Physics}, 129(22), 2021.

	\bibitem{IE10}
	Yihuang Xiong, C{\'e}line Bourgois, Natalya Sheremetyeva, Wei Chen, Diana Dahliah, Hanbin Song, Jiongzhi Zheng, Sin{\'e}ad~M Griffin, Alp Sipahigil, and Geoffroy Hautier.
	\newblock High-throughput identification of spin-photon interfaces in silicon.
	\newblock {\em Science Advances}, 9(40):eadh8617, 2023.

	\bibitem{lu2023explainable}
	Grace~M Lu, Matthew Witman, Sapan Agarwal, Vitalie Stavila, and Dallas~R Trinkle.
	\newblock Explainable machine learning for hydrogen diffusion in metals and random binary alloys.
	\newblock {\em Physical Review Materials}, 7(10):105402, 2023.

	\bibitem{lopanitsyna2023modeling}
	Nataliya Lopanitsyna, Guillaume Fraux, Maximilian~A Springer, Sandip De, and Michele Ceriotti.
	\newblock Modeling high-entropy transition metal alloys with alchemical compression.
	\newblock {\em Physical Review Materials}, 7(4):045802, 2023.

	\bibitem{lu2025explainable}
	Grace~M Lu and Dallas~R Trinkle.
	\newblock Explainable machine learning for oxygen diffusion in perovskites and pyrochlores.
	\newblock {\em Physical Review Materials}, 9(11):115402, 2025.

	\bibitem{fiedler2022deep}
	Lenz Fiedler, Karan Shah, Michael Bussmann, and Attila Cangi.
	\newblock Deep dive into machine learning density functional theory for materials science and chemistry.
	\newblock {\em Physical Review Materials}, 6(4):040301, 2022.

	\bibitem{I16}
	Chen Li and Kun Zheng.
	\newblock Methods, progresses, and opportunities of materials informatics.
	\newblock {\em InfoMat}, 5(8):e12425, 2023.

	\bibitem{I18}
	Oliver~Lerst{\o}l Hebnes, Marianne~Etzelm{\"u}ller Bathen, {\O}yvind~Sigmundson Sch{\o}yen, Sebastian~G Winther-Larsen, Lasse Vines, and Morten Hjorth-Jensen.
	\newblock Predicting solid state material platforms for quantum technologies.
	\newblock {\em npj Computational Materials}, 8(1):207, 2022.

	\bibitem{rudin_stop_2019}
	Cynthia Rudin.
	\newblock Stop explaining black box machine learning models for high stakes decisions and use interpretable models instead.
	\newblock {\em Nature Machine Intelligence}, 1:206--215, 2019.

	\bibitem{molnar_2020}
	Christoph Molnar.
	\newblock {\em Interpretable Machine Learning}.
	\newblock Lulu. com, 3 edition, 2025.

	\bibitem{MP1}
	Anubhav Jain, Shyue~Ping Ong, Geoffroy Hautier, Wei Chen, William~Davidson Richards, Stephen Dacek, Shreyas Cholia, Dan Gunter, David Skinner, Gerbrand Ceder, et~al.
	\newblock Commentary: The materials project: A materials genome approach to accelerating materials innovation.
	\newblock {\em APL materials}, 1(1), 2013.

	\bibitem{icsd1}
	Dejan Zagorac, H~M{\"u}ller, S~Ruehl, J~Zagorac, and Silke Rehme.
	\newblock Recent developments in the inorganic crystal structure database: theoretical crystal structure data and related features.
	\newblock {\em Journal of applied crystallography}, 52(5):918--925, 2019.

	\bibitem{matminer}
	Logan Ward, Alexander Dunn, Alireza Faghaninia, Nils~ER Zimmermann, Saurabh Bajaj, Qi~Wang, Joseph Montoya, Jiming Chen, Kyle Bystrom, Maxwell Dylla, et~al.
	\newblock Matminer: An open source toolkit for materials data mining.
	\newblock {\em Computational Materials Science}, 152:60--69, 2018.

	\bibitem{MFE1}
	Logan Ward, Ankit Agrawal, Alok Choudhary, and Christopher Wolverton.
	\newblock A general-purpose machine learning framework for predicting properties of inorganic materials.
	\newblock {\em npj Computational Materials}, 2(1):1--7, 2016.

	\bibitem{MFE2}
	Bryce Meredig, Ankit Agrawal, Scott Kirklin, James~E Saal, Jeff~W Doak, Alan Thompson, Kunpeng Zhang, Alok Choudhary, and Christopher Wolverton.
	\newblock Combinatorial screening for new materials in unconstrained composition space with machine learning.
	\newblock {\em Physical Review B}, 89(9):094104, 2014.

	\bibitem{ST}
	Logan Ward, Ankit Agrawal, Alok Choudhary, and Christopher Wolverton.
	\newblock A general-purpose machine learning framework for predicting properties of inorganic materials.
	\newblock {\em npj Computational Materials}, 2(1):1--7, 2016.

	\bibitem{AO}
	Svetlana Kotochigova, Zachary~H Levine, Eric~L Shirley, Mark~D Stiles, and Charles~W Clark.
	\newblock Local-density-functional calculations of the energy of atoms.
	\newblock {\em Physical Review A}, 55(1):191, 1997.

	\bibitem{BC}
	M~A Butler and D~S Ginley.
	\newblock {Prediction of Flatband Potentials at Semiconductor-Electrolyte Interfaces from Atomic Electronegativities}.
	\newblock {\em Journal of The Electrochemical Society}, 125(2):228--232, feb 1978.

	\bibitem{Apley2020}
	Daniel~W. Apley and Jingyu Zhu.
	\newblock Visualizing the effects of predictor variables in black box supervised learning models.
	\newblock {\em Journal of the Royal Statistical Society Series B: Statistical Methodology}, 82(4):1059--1086, 06 2020.

	\bibitem{M4}
	Davide Chicco and Giuseppe Jurman.
	\newblock The matthews correlation coefficient (mcc) should replace the roc auc as the standard metric for assessing binary classification.
	\newblock {\em BioData Mining}, 16(1):4, 2023.

	\bibitem{rashomon}
	Leo Breiman.
	\newblock Statistical modeling: The two cultures.
	\newblock {\em Quality control and applied statistics}, 48(1):81--82, 2003.

	\bibitem{rashomon1}
	Sichao Li, Amanda~S Barnard, and Quanling Deng.
	\newblock Practical attribution guidance for rashomon sets.
	\newblock {\em arXiv preprint arXiv:2407.18482}, 2024.

	\bibitem{MFE4}
	Maryam Sabzevari, Gonzalo Mart{\'\i}nez-Mu{\~n}oz, and Alberto Su{\'a}rez.
	\newblock Building heterogeneous ensembles by pooling homogeneous ensembles.
	\newblock {\em International Journal of Machine Learning and Cybernetics}, pages 1--8, 2022.

	\bibitem{rashomon2}
	Jiayun Dong and Cynthia Rudin.
	\newblock Exploring the cloud of variable importance for the set of all good models.
	\newblock {\em Nature Machine Intelligence}, 2(12):810--824, 2020.

	\bibitem{rashomon3}
	Katarzyna Kobyli{\'n}ska, Mateusz Krzyzi{\'n}ski, Rafa{\l} Machowicz, Mariusz Adamek, and Przemys{\l}aw Biecek.
	\newblock Exploration of the rashomon set assists trustworthy explanations for medical data.
	\newblock {\em IEEE Journal of Biomedical and Health Informatics}, 28(11):6454--6465, 2024.

	\bibitem{fisher_2019}
	Aaron Fisher, Cynthia Rudin, and Francesca Dominici.
	\newblock All models are wrong, but many are useful: Learning a variable's importance by studying an entire class of prediction models simultaneously.
	\newblock {\em Journal of Machine Learning Research}, 20(177):1--81, 2019.

	\bibitem{sklearn}
	F.~Pedregosa, G.~Varoquaux, A.~Gramfort, V.~Michel, B.~Thirion, O.~Grisel, M.~Blondel, P.~Prettenhofer, R.~Weiss, V.~Dubourg, J.~Vanderplas, A.~Passos, D.~Cournapeau, M.~Brucher, M.~Perrot, and E.~Duchesnay.
	\newblock Scikit-learn: Machine learning in {P}ython.
	\newblock {\em Journal of Machine Learning Research}, 12:2825--2830, 2011.

	\bibitem{chicco2023matthews}
	Davide Chicco and Giuseppe Jurman.
	\newblock The matthews correlation coefficient (mcc) should replace the roc auc as the standard metric for assessing binary classification.
	\newblock {\em BioData Mining}, 16(1):4, 2023.

	\bibitem{friedman2001greedy}
	Jerome~H Friedman.
	\newblock Greedy function approximation: a gradient boosting machine.
	\newblock {\em Annals of statistics}, pages 1189--1232, 2001.

	\bibitem{goldstein2015peeking}
	Alex Goldstein, Adam Kapelner, Justin Bleich, and Emil Pitkin.
	\newblock Peeking inside the black box: Visualizing statistical learning with plots of individual conditional expectation.
	\newblock {\em journal of Computational and Graphical Statistics}, 24(1):44--65, 2015.

	\bibitem{apley2020visualizing}
	Daniel~W Apley and Jingyu Zhu.
	\newblock Visualizing the effects of predictor variables in black box supervised learning models.
	\newblock {\em Journal of the Royal Statistical Society Series B: Statistical Methodology}, 82(4):1059--1086, 2020.

	\bibitem{lundberg2017unified}
	Scott~M Lundberg and Su-In Lee.
	\newblock A unified approach to interpreting model predictions.
	\newblock {\em Advances in neural information processing systems}, 30, 2017.

	\bibitem{alibi}
	Janis Klaise, Arnaud Van~Looveren, Giovanni Vacanti, and Alexandru Coca.
	\newblock Alibi explain: Algorithms for explaining machine learning models.
	\newblock {\em Journal of Machine Learning Research}, 22(181):1--7, 2021.

	\bibitem{Kresse1996}
	G.~Kresse and J.~Furthm{\"u}ller.
	\newblock Efficient iterative schemes for ab initio total-energy calculations using a plane-wave basis set.
	\newblock {\em Phys. Rev. B}, 54(16):11169--11186, 1996.

	\bibitem{Kresse1999}
	G.~Kresse and D.~Joubert.
	\newblock From ultrasoft pseudopotentials to the projector augmented-wave method.
	\newblock {\em Phys. Rev. B}, 59(3):1758--1775, 1999.

	\bibitem{Blochl1994}
	P.~E. Bl{\"o}chl.
	\newblock Projector augmented-wave method.
	\newblock {\em Phys. Rev. B}, 50(24):17953--17979, 1994.

	\bibitem{Perdew1996}
	J.~P. Perdew, K.~Burke, and M.~Ernzerhof.
	\newblock Generalized gradient approximation made simple.
	\newblock {\em Phys. Rev. Lett.}, 77(18):3865--3868, 1996.

	\bibitem{Freysoldt2009}
	C.~Freysoldt, J.~Neugebauer, and C.~G.~Van de~Walle.
	\newblock Fully ab initio finite-size corrections for charged-defect supercell calculations.
	\newblock {\em Phys. Rev. Lett.}, 102(1):016402, 2009.

	\bibitem{Zhang1991}
	S.~B. Zhang and J.~E. Northrup.
	\newblock Chemical potential dependence of defect formation energies in gaas: Application to ga self-diffusion.
	\newblock {\em Phys. Rev. Lett.}, 67(17):2339--2342, 1991.

	\bibitem{Lany2008}
	S.~Lany and A.~Zunger.
	\newblock Assessment of correction methods for the band-gap problem and for finite-size effects in supercell defect calculations: Case studies for zno and gaas.
	\newblock {\em Phys. Rev. B}, 78(23):235104, 2008.

	\bibitem{Baroni_1986}
	Stefano Baroni and Raffaele Resta.
	\newblock Ab initio calculation of the macroscopic dielectric constant in silicon.
	\newblock {\em Phys. Rev. B}, 33:7017--7021, 05 1986.

	\bibitem{DFPT_2006}
	M.~Gajdo\ifmmode~\check{s}\else \v{s}\fi{}, K.~Hummer, G.~Kresse, J.~Furthm\"uller, and F.~Bechstedt.
	\newblock Linear optical properties in the projector-augmented wave methodology.
	\newblock {\em Phys. Rev. B}, 73:045112, 01 2006.

	\bibitem{mgo}
	Vrindaa Somjit, Joel Davidsson, Yu~Jin, and Giulia Galli.
	\newblock An nv- center in magnesium oxide as a spin qubit for hybrid quantum technologies.
	\newblock {\em npj Computational Materials}, 11(1):74, 2025.

	\bibitem{kanai}
	Shun Kanai, F~Joseph Heremans, Hosung Seo, Gary Wolfowicz, Christopher~P Anderson, Sean~E Sullivan, Mykyta Onizhuk, Giulia Galli, David~D Awschalom, and Hideo Ohno.
	\newblock Generalized scaling of spin qubit coherence in over 12,000 host materials.
	\newblock {\em Proceedings of the National Academy of Sciences}, 119(15):e2121808119, 2022.

	\bibitem{M1MgO}
	Vrindaa Somjit, Joel Davidsson, Yu~Jin, and Giulia Galli.
	\newblock An nv- center in magnesium oxide as a spin qubit for hybrid quantum technologies.
	\newblock {\em npj Computational Materials}, 11(1):74, 2025.

	\bibitem{qian2025first}
	Tangjiang Qian, Xin-Gao Gong, and Ji-Hui Yang.
	\newblock First-principles studies of hydrogen irradiation effects on the photoluminescence properties of nitrogen-vacancy centers in 4h-sic.
	\newblock {\em Physical Review Materials}, 9(11):116201, 2025.

	\bibitem{M1ZnO}
	Vasileios Niaouris, Mikhail~V Durnev, Xiayu Linpeng, Maria~LK Viitaniemi, Christian Zimmermann, Aswin Vishnuradhan, Yusuke Kozuka, Masashi Kawasaki, and Kai-Mei~C Fu.
	\newblock Ensemble spin relaxation of shallow donor qubits in zno.
	\newblock {\em Physical Review B}, 105(19):195202, 2022.

	\bibitem{hoang2022rare}
	Khang Hoang.
	\newblock Rare-earth defects and defect-related luminescence in zns.
	\newblock {\em Journal of Applied Physics}, 131(1), 2022.

	\bibitem{M1TMD}
	Yeonghun Lee, Yaoqiao Hu, Xiuyao Lang, Dongwook Kim, Kejun Li, Yuan Ping, Kai-Mei~C Fu, and Kyeongjae Cho.
	\newblock Spin-defect qubits in two-dimensional transition metal dichalcogenides operating at telecom wavelengths.
	\newblock {\em Nature Communications}, 13(1):7501, 2022.

	\bibitem{lang2025solid}
	Xiuyao Lang, Matthew Bergschneider, and Kyeongjae Cho.
	\newblock Solid-state quantum defects in wide band-gap two-dimensional silica bilayer.
	\newblock {\em Physical Review Materials}, 9(2):026201, 2025.

	\bibitem{ong2015materials}
	Shyue~Ping Ong, Shreyas Cholia, Anubhav Jain, Miriam Brafman, Dan Gunter, Gerbrand Ceder, and Kristin~A Persson.
	\newblock The materials application programming interface (api): A simple, flexible and efficient api for materials data based on representational state transfer (rest) principles.
	\newblock {\em Computational Materials Science}, 97:209--215, 2015.

	\bibitem{nagura2026understanding}
	Jonah Nagura, Mykyta Onizhuk, and Giulia Galli.
	\newblock Understanding surface-induced decoherence of nv centers in diamond.
	\newblock {\em Physical Review Materials}, 10(2):024603, 2026.

	\bibitem{C2}
	Joel Davidsson, Mykyta Onizhuk, Christian Vorwerk, and Giulia Galli.
	\newblock Discovery of atomic clock-like spin defects in simple oxides from first principles.
	\newblock {\em Nature Communications}, 15(1):4812, 2024.

	\bibitem{jeong2022strong}
	Heonjae Jeong and Edmund~G Seebauer.
	\newblock Strong isotopic fractionation of oxygen in tio2 obtained by surface-enhanced solid-state diffusion.
	\newblock {\em The Journal of Physical Chemistry Letters}, 13(42):9841--9847, 2022.

	\bibitem{gilliard2017manipulation}
	Kandis~Leslie Gilliard and Edmund~G Seebauer.
	\newblock Manipulation of native point defect behavior in rutile tio2 via surfaces and extended defects.
	\newblock {\em Journal of Physics: Condensed Matter}, 29(44):445002, 2017.

	\bibitem{MFE5}
	Mohammad Alghadeer, Nufida~D Aisyah, Mahmoud Hezam, Saad~M Alqahtani, Ahmer~AB Baloch, and Fahhad~H Alharbi.
	\newblock Machine learning prediction of materials properties from chemical composition: Status and prospects.
	\newblock {\em Chemical Physics Reviews}, 5(4), 2024.

	\bibitem{mahshook2025}
	Mohammed Mahshook and Rudra Banerjee.
	\newblock Data-hosting repository for beyond diamond: Interpretable machine learning reveals design principles for quantum defect host materials.
	\newblock \url{https://github.com/Mahshook-Q/NV_ml_data}, 2026.

\end{thebibliography}

\end{document}